\documentclass[twocolumn]{aastex631}

\usepackage{bm}
\usepackage{amssymb}
\usepackage{amsmath}
\usepackage{mathtools}
\usepackage{gensymb}

\renewcommand{\vec}{\bm}
\newcommand{\mat}{\mathbf}
\DeclareMathOperator\erf{erf}
\newcommand{\rmd}{{\rm d}}

\renewcommand{\refeq}[1]{Eq.~(\ref{eq:#1})}          
          
\newcommand{\reffig}[1]{Fig.~\ref{fig:#1}}          
\newcommand{\refsec}[1]{Sec.~\ref{sec:#1}}
\newcommand{\refapp}[1]{App.~\ref{app:#1}}

\begin{document}

\title{Effects of Subhalos on Interpreting Highly Magnified Sources Near Lensing Caustics}

\author[0000-0002-3947-7362]{Lingyuan Ji}
\affiliation{Berkeley Center for Cosmological Physics, Campbell Hall 341 University of California, Berkeley, CA 94720, USA}
\affiliation{Department of Physics, University of California, 366 Physics North MC 7300, Berkeley, CA. 94720, USA}

\author[0000-0003-2091-8946]{Liang Dai}
\affiliation{Department of Physics, University of California, 366 Physics North MC 7300, Berkeley, CA. 94720, USA}

\correspondingauthor{Liang Dai}
\email{liangdai@berkeley.edu}

\shorttitle{Subhalos and highly magnified sources}
\shortauthors{Ji and Dai}

\begin{abstract}

    Large magnification factors near gravitational lensing caustics of galaxy cluster lenses allow the study of individual stars or compact stellar associations at cosmological distances. We study how the presence of sub-galactic subhalos, an inevitable consequence of cold dark matter, can alter the property of caustics and hence change the interpretation of highly magnified sources that lie atop them. First, we consider a galaxy cluster halo populated with subhalos sampled from a realistic subhalo mass function calibrated to $N$-body simulations. Then, we compare a semi-analytical approximation and an adaptive ray-shooting method which we employ to quantify the property of the caustics. As a case study, we investigate Earendel, a $z = 6.2$ candidate of magnified single or multiple star system with a lone lensed image atop the critical curve in the Sunrise Arc. We find that the source size constraint ($\lesssim 0.3\, \mathrm{pc}$) previously derived from macro lens models should be relaxed by a factor of a few to ten when subhalos are accounted for, therefore allowing the possibility of a compact star cluster. The subhalos could introduce an astrometric perturbation that is $\lesssim 0.5''$, which does not contradict observation. These conclusions are largely robust to changes in the subhalo population. Subhalos therefore should be seriously accounted for when interpreting the astrophysical nature of similar highly magnified sources uncovered in recent high-$z$ observations.
    
\end{abstract}

\section{Introduction}

Gravitational magnification of light provides the best opportunities for the study of individual stars or stellar associations at cosmological distances. As first suggested by \cite{Miralda1991CausticCrossingStars}, individual luminous stars are detectable owing to large lensing magnification factors $\mu\sim \mathcal{O}(10^2$--$10^4)$ when they are fortuitously located in the vicinity of caustic curves cast by foreground massive lenses. The extreme magnification values realized there make highly magnified stars sensitive probes of dark matter substructure on scales comparable or much smaller than that of dwarf galaxies~\citep{Dai2018Abell370, Dai2020SGASJ1226, Dai2020QCDAxionMinihalos, Williams2023FlashlightsDMSubhalo, Diego2024arXiv240408033D, Broadhurst2024arXiv240519422B} and of compact objects which may make up a fraction of the dark matter~\citep{Diego2018DMUnderMicroscope, Oguri2018CausticMicrolensing, Mueller2024arXiv240316989V}. They may also offer excellent cases for studying the properties of massive stars in star formation sites at cosmological redshifts through multi-band photometry~\citep{Han2024HighlyMagnifiedStar} or spectroscopy~\citep{Lundqvist2024arXiv240410817L}, or the accreting remnants of massive stars~\citep{Windhorst2018PopIIILensedStars}.

Indeed, individual stars subject to extremely large magnification factors have thus far been detected behind galaxy cluster lenses, where it is convenient that the sightline toward the host galaxy is not severely contaminated by the diffuse starlight of cluster member galaxies. Several earlier detections from $z \sim 0.7$--$1.5$ exhibited time-varying fluxes over the timescale of days to years~\citep{Kelly2018NatAsM1149, Rodney2018NatAsM0416Transients, Chen2019LensedStarM0416, Kaurov2019LensedStarM0416}, which can be explained in terms of the macro caustic disrupted into a network of micro caustics cast by foreground intracluster stars~\citep{Venumadhav2017CausticMicrolensing, Diego2018DMUnderMicroscope, Oguri2018CausticMicrolensing} (also see \cite{Weisenbach2024MLnearMacroCausticsReview} for a topical review). Taking advantage of microlensing-induced variability, recent studies have demonstrated that highly magnified individual stars can be efficiently detected in caustic-crossing host galaxies at these redshifts through image difference of multi-epoch deep exposures, with the Hubble Space Telescope (HST)~\citep{Kelly2022FlashlightDozenStars, Meena2023Abell370LensedStar} and with the James Webb Space Telescope (JWST)~\citep{Yan2023PEARLStransients, Fudamoto2024Abell370Dragon}. 

Intriguingly, recent observations have also uncovered additional lensed-star candidates in higher-$z$ galaxies that show different characteristics. While these sources appear close to macro caustic curves and hence must be highly magnified, they do not show a resolved pair of lensed images as expected near a fold caustic and neither does the lone image exhibit detectable flux variability. One notable example is Earendel in the $z=6.2$ Sunrise galaxy, which has been suggested to be a single star or a system of binary or multiple stars~\citep{Welch2022EarendelNature, Welch2022EarendelJWSTimaging}. Another exceptionally bright lensed source, nicknamed Godzilla and located in the $z=2.37$ Sunburst galaxy, appears to be a single stellar object as it also shows as a lone image without variability~\citep{Vanzella2020Bowen, Diego2022Godzilla}. \cite{Choe2024Godzilla} propose that the source is a post-eruption luminous blue variable, while \cite{Pascale2024Godzilla} instead suggest that it is a young star cluster with a large number of hot stars powering nebular emission. \cite{Furtak2023arXiv230800042F} performed JWST/NIRSpec spectroscopy of another highly magnified compact source at $z=4.7$ with a lone, persistent lensed image on the critical curve, and suggested that the source is likely a binary star due to a source-size constraint $\lesssim 0.5\,$pc from lens modeling. If any of the above is truly an individual stellar object, or even proven to be a metal poor or zero-metallicity star~\citep{Schauer2022EarendelPopIII, Zackrisson2023PopIII}, it will be a remarkable discovery about stellar objects in the young Universe

In \cite{Welch2022EarendelNature}, one key reasoning is that the lone, unresolved image of Earendel right on top of the theoretically predicted lensing critical curve requires that the source must be more compact than $\sim 0.3\,$pc in radius. The authors conclude that Earendel cannot be a star cluster but must be a single star, or a small system of binary or multiple stars~\citep{Welch2022EarendelJWSTimaging}. The tight size constraint derives from very small values for the third-order derivatives of the Fermat potential on the critical curve, which is a consequence of the large Einstein angular scale for the galaxy cluster lens; in other words, the fold caustic associated with the galaxy cluster lens has a high strength~\citep{SchneiderEhlersFalco1992textbook}. This small third-order derivative is independently inferred by several macro lens models constructed for Earendel's cluster lensing field, with a factor of several uncertainty between models~\citep{Welch2022EarendelNature}. While the non-detection of large temporal flux change would be natural for a lensed star cluster with many member stars~\citep{Dai2021StarClusterMicrolensing}, for a single stellar source the absence could be the consequence of a hyper-critical density of micro caustics due to extremely large macro magnification factors for a finite-size source~\citep{Dai2021NewApproxMagnificationStatistics} (this argument also put forth in \cite{Furtak2023arXiv230800042F}).

Here we point out that the large caustic strength is inferred only from a smooth macro lens model --- one that only accounts for dark matter subhalos massive enough to host optically visible galaxies. If the dark matter is cold, hierarchical structure formation should lead to a continuous spectrum of subhalo masses~\citep{Press1974PressSchechter} that may extend well below the galactic mass scale ($\lesssim 10^{10}\,M_\odot$). At cosmological distances, the sub-galactic DM halos that reside within the foreground galaxy cluster lens are not expected to host optically detectable stellar populations~\citep{Okamoto2008StarFormationInefficiencySmallGalaxies}. The perturbing effects from sub-galactic subhalos become prominent near a macro critical curve~\citep{Dai2018Abell370, Dai2020SGASJ1226, Williams2023FlashlightsDMSubhalo}. While these sub-galactic subhalos make tiny corrections to ray deflection, convergence and shear, they may increase the third-order derivatives of the Fermat potential and hence decrease the typical caustic strengths. In this paper, we will show that this effect significantly relaxes the size constraint for Earendel so that it may as well be a compact star cluster several parsecs across. In general, we suggest that the effect of subhalos should be taken into account when interpreting highly magnified sources similar to Earendel.

This paper is organized as follows. In \refsec{lensmodel}, we start with a review of lensing in the vicinity of fold caustics and discuss the semi-analytic model we adopt to model subhalos as lensing perturbers. Then in \refsec{observables}, we rigorously define observables that we can use to quantify the constraint on the size of the source as well as the position of the photometric center. In \refsec{methods}, we introduce two methods which we adopt to calculate those observables for numerically simulated lensing realizations with dark matter subhalos. Results will be presented and discussed in \refsec{results}. The main body of this paper will end in \refsec{concl} with concluding remarks. 

We additionally collect technical details related to this work in two appendices. In \refapp{detailed-lens-models} the readers can find details on our treatment of lensing by the macro fold caustic and by the dark matter subhalos. In \refapp{detailed-size-constraints} we present an analytic derivation of the constraint on the source size in the vicinity of a fold caustic.

\section{Lens Model}
\label{sec:lensmodel}

As an important case study, we will analyze Earendel as reported in \citet{Welch2022EarendelNature}. The host galaxy WHL0137-zD1 at $z_\mathrm{S} = 6.2$ is lensed by the foreground galaxy cluster WHL0137-08 at $z_\mathrm{L} = 0.566$, forming a $15''$-long Sunrise Arc~\citep{Salmon2020RELICSsurvey}. Earendel is located almost exactly on the macro model critical curve and was suspected to be highly magnified.

We approximate the lens mapping in the vicinity of Earendel by accounting for three components of lenses:  (i) a smooth fold caustic~\citep{SchneiderEhlersFalco1992textbook} representing the effect of the coarse-grained mass distribution of the cluster halo as if mass locked up in sub-galactic subhalos is all smoothed out; (ii) a collection of subhalos assumed to have a tidally truncated Navarro--Frenk--White (NFW) profile~\citep{Navarro1996NFWprofile} with random center positions within a sufficiently large circular region on the image plane centered on Earendel; (iii) a (negative) constant surface mass density added in the same circular region to enforce mass conservation after adding subhalos. 

Let $\vec \alpha_\mathrm{FD}(\vec x)$, $\vec \alpha_\mathrm{SH}^{(n)}(\vec x)$, and $\vec \alpha_\mathrm{UD}(\vec x)$ be the ray deflection vector of the smooth local fold model, that from the $n$-th subhalo, and that from the uniform negative-mass disk, respectively. The total deflection vector is
\begin{equation}
    \vec \alpha (\vec x) = \vec \alpha_\mathrm{FD}(\vec x) + \sum_{n = 1}^N \vec \alpha_\mathrm{SH}^{(n)}(\vec x) + \vec \alpha_\mathrm{UD}(\vec x),
\end{equation}
which enters the ray equation
\begin{equation}\label{eq:ray-equation}
    \vec y = \vec x - \vec \alpha(\vec x).
\end{equation}
In what follows, we briefly describe the implemention of components (i) and (ii), while component (iii) is easy to treat. More detailed information for all components can be found in \refapp{detailed-lens-models}.

\subsection{Smooth Fold}
\label{sec:smooth-fold}

Near the critical curve, the Cartesian components of the deflection vector caused by the smooth fold can be expressed as a power expansion
\begin{equation}
    \alpha_{i, \mathrm{FD}}(\vec x) = x_i - \phi^{\vec 0}_{ij}\,x_j - \frac{1}{2}\, \phi^{\vec 0}_{ijk}\,x_j\,x_k + O(|\vec x|^3).
\end{equation}
Here, $\phi(\vec x, \vec y) \equiv (\vec x - \vec y)^2 / 2 - \psi(\vec x)$ is the Fermat potential with $\psi(\vec x)$ being the lensing potential. We adopt the short-hand notation $\phi_{ij \cdots} \equiv \partial \phi / (\partial x_i \partial x_j \cdots)$ for its derivatives at various orders. The bold-face superscript $\vec 0$ indicates that the quantity is evaluated at the coordinate origin $\vec x = \vec 0$, which is taken to be the location of Earendel. The first and second derivatives of deflection can be found in \refapp{fold-catastrophe}. Since we concern the vicinity of the smooth critical curve, $|\vec x|$ is small and hence from now on we keep only the leading order term.

The smooth fold determines the main features of the criticality. If the smooth fold is the \emph{only} component of the lens model, the Jacobian matrix $A_{ij} \equiv \partial y_i / \partial x_j = \phi_{ij}$ has two (normalized) eigenvectors $(\vec e_\parallel, \vec e_\perp)$, associated with two eigenvalues $(\lambda_\parallel, \lambda_\perp)$, respectively. The eigenvectors are given by
\begin{equation}
    \vec e_\parallel = \vec e ^{\vec 0}_\parallel \quad \text{and} \quad \vec e_\perp = \vec e ^{\vec 0}_\perp,
\end{equation}
where $\vec e ^{\vec 0}_\parallel$ and $\vec e ^{\vec 0}_\perp$ are the eigenvectors associated with the zero and non-zero eigenvalues of $A^{\vec 0}_{ij} = \phi^{\vec 0}_{ij}$, respectively.  The eigenvalues are given by
\begin{equation}
    \lambda_\parallel = \vec d^{\vec 0} \cdot \vec x \quad \text{and} \quad \lambda_\perp = 2\,(1-\kappa^{\vec 0}),
\end{equation}
where $\kappa^{\vec 0} \equiv 1 - (\phi^{\vec 0}_{11} + \phi^{\vec 0}_{22}) / 2$ and $\vec d^{\vec 0} \equiv \vec D^{\vec 0} / [2\,(1-\kappa^{\vec 0})]$. Here, we define the Jacobian gradient vector $\vec D^{\vec 0}$ which has Cartesian components
\begin{equation}
    D^{\vec 0}_i \equiv \phi^{\vec 0}_{22}\,\phi^{\vec 0}_{11i} + \phi^{\vec 0}_{11}\,\phi^{\vec 0}_{22i} - 2\,\phi^{\vec 0}_{12}\,\phi^{\vec 0}_{12i}.
\end{equation}
We term $\vec e_\parallel$ the degenerate direction (i.e. the elongation direction of the lensed arc) and $\vec e_\perp$ the perpendicular direction.  We define 
\begin{equation}\label{eq:smooth-fold-mu_para-mu_perp}
    \mu_\parallel \equiv \frac{1}{\lambda_\parallel} = \frac{1}{\vec d ^{\vec 0} \cdot \vec x} \quad \text{and} \quad \mu_\perp \equiv \frac{1}{\lambda_\perp} = \frac{1}{2 \,(1-\kappa^{\vec 0})}
\end{equation}
to be the magnification in the degenerate direction and the perpendicular direction, respectively. Thus, $\mu_\parallel$ has a large absolute value that increases toward the smooth critical curve, while $\mu_\perp$ has a constant moderate value. The determinant of the Jacobian matrix $A_{ij}(\vec x) \equiv \phi_{ij}(\vec x)$ is
\begin{equation}
    \det A(\vec x) = \lambda_\parallel\,\lambda_\perp = \vec D^{\vec 0} \cdot \vec x,
\end{equation}
which implies that $\vec D^{\vec 0}$ is normal to the critical curve and the flux magnification is
\begin{equation}
    \mu = \mu_\parallel\,\mu_\perp = \frac{1}{\vec D^{\vec 0} \cdot \vec x}.
\end{equation}

The magnitude $D^{\vec 0} \equiv |\vec D^{\vec 0}|$ of the vector $\vec D^{\vec 0}$ therefore is the combination of the third-order derivatives that quantifies the strength of the caustic as it determines the magnification factor of a lensed image at given distance to the critical curve. A stronger caustic has a smaller $D^{\vec 0}$ value, so that a lensed image at a fixed distance from the critical curve has a larger magnification factor than in the case of a weaker caustic with a larger $D^{\vec 0}$ value.

Several macro lens models have been constructed for the Sunrise Arc without including the numerous subhalos that are not expected to host visible galaxies. These models make moderately different predictions for the criticality near Earendel~\citep{Welch2022EarendelNature}. At the location of Earendel, lens-model predictions for the macro convergence fall in the range $\kappa^{\vec 0} = 0.5$--$0.8$, and we shall use a fiducial value $\kappa^{\vec 0} = 0.5$. The degenerate direction and the critical direction (i.e. the tangential direction of the critical curve at the location of Earendel) form an angle $\alpha$ whose value is in the range $22$--$41 \degree$, and we shall use a fiducial value $\alpha = 30 \degree$. The Jacobian gradient $\vec D^{\vec 0}$ of the smooth fold is perpendicular to the critical curve with a model-predicted magnitude in the range $D^{\vec 0}= 1/(113'')$--$1 / (18'')$. We will numerically study the case $D^{\vec 0} = 1/(18'')$. There are two arguments for this choice.  First, as we will show in \refsec{size-constraints}, the largest $D^{\vec 0}$ implies the weakest caustic and thus the most conservative size constraint for Earendel.  This most conservative case is crucial for \citet{Welch2022EarendelNature} to rule out the star cluster scenario. Second, the weakest caustic is also the one that is most immune to perturbations from subhalos, due to their already large third-order derivatives. As a result, subhalos will be important to all other possible $D^{\vec 0}$ values in the range $D^{\vec 0}= 1/(113'')$--$1 / (18'')$ if it is important for the case where $D^{\vec 0} = 1/(18'')$.

\subsection{Subhalo Abundance}
\label{sec:subhalo-abundance}

Direct $N$-body simulations of a galaxy cluster halo that resolve subhalos with sub-galactic masses $m=10^6$--$10^{10}\,M_\odot$ is computationally demanding. Instead, we adopt a semi-analytical model devised by \citet{Han:2015pua} that is calibrated to and verified by simulations for the more massive subhalos and then extrapolated to sub-galactic subhalo masses. This detailed model accounts for the effects of subhalo infall and tidal stripping. The essential parameters of the model are the host-halo density profile $\rho(R)$ as a function of the host-halo-centric three-dimensional radius $R$ and the lower (upper) limit of the subhalo infall mass $m_\mathrm{min}$ ($m_\mathrm{max}$). The model outputs a joint distribution $\rmd^2 N / (\rmd \ln m\, \rmd^3 R)$ as a function of the subhalo mass $m$ and the location $R$. The explicit expression for the distribution, as well as other secondary parameters, are provided in \refapp{subhalo-abundance}. 

We then map the three-dimensional spatial distribution to the projected two-dimensional distribution needed for studying the lensing effect, through a change of variable
\begin{equation}
    \frac{\rmd N(m, Z, B)}{\rmd \ln m \, \rmd Z\, \rmd^2 B} = \frac{\rmd N(m, \sqrt{Z^2 + B^2})}{\rmd \ln m\, \rmd^3 R},
\end{equation}
where $B$ is the host-halo-centric impact parameter and $Z$ is the longitudinal coordinate, related by $R^2 = Z^2 + B^2$. In the vicinity of Earendel, the variation in $B$ is negligible. Thus, for a given impact parameter $B$ and within a field of view of area $\Delta S$, we draw subhalos from the distribution
\begin{equation} \label{eq:subhalo-distribution-2d}
    \left. \frac{\rmd N(m, Z)}{\rmd \ln m \, \rmd Z} \right|_B = \frac{\rmd N(m, Z, B)}{\rmd \ln m \, \rmd Z \, \rmd^2 B} \times \Delta S,
\end{equation}
and randomly distribute them within the field of view. It is important that we sample the joint distribution of $(m, Z)$, as the subhalo parameters $(c_{200}, r_\mathrm{t})$ at given $m$ depends on the three-dimensional radial distance $R = (Z^2 + B^2)^{1/2}$.

The host halo of the galaxy cluster WHL0137-08 has an estimated mass $M_{500} = 9 \times 10^{14}\,M_\odot$ with uncertainty in the concentration parameter $C_{500}$ \citep{Welch2022EarendelNature}. We assume that the host halo can be described by the Navarro-Frenk-White (NFW) model and follow \citet{2019ApJ...871..168D} to obtain $C_{500} = 2.6$. We choose a mass $M_{200} = 1.3\times 10^{15}\, M_\odot$ and a concentration $C_{200} = 4.0$ for the host halo at the density contrast $200$. These numbers decently reproduce the NFW parameter values at the density contrast 500. The line of sight toward Earendel is roughly $40''$ away from the brightest cluster galaxy, which we take to be the center of the host halo. The impact parameter with respect to the host halo center is $B = 268\, \mathrm{kpc}$. Since we mainly concern the effect of the more numerous sub-galactic subhalos, we shall fix $m_\mathrm{max} = 10^{9} M_\odot$, but explore different $m_\mathrm{min}$ values.

\subsection{Subhalo Density Structure}

Following \cite{Dai2018Abell370}, we model the subhalos using the truncated NFW profile~\citep{Baltz2009AnalyticLensModels, CyrRacine2016DarkCensus, Dai2018Abell370}
\begin{equation}
    \rho_{\rm SH}(r) = \frac{m_{200}}{4\pi\,r_\mathrm{s}^3\, f(c_{200})} \frac{1}{(r/r_\mathrm{s})(1+r/r_\mathrm{s})^2} \frac{1}{1+r^2/r_\mathrm{t}^2},
\end{equation}
where $f(x) \equiv \ln(1+x) - x/(1+x)$.  This model is parameterized by 3 independent parameters: the underlying NFW mass $m_{200}$, the underlying NFW concentration parameter $c_{200}$, and a tidal truncation radius $r_\mathrm{t}$. The underlying NFW scale radius is then given in terms of $m_{200}$ and $c_{200}$
\begin{equation}
    r_\mathrm{s} \equiv \frac{1}{c_{200}}\left(\frac{3}{4\pi} \frac{m_{200}}{200\,\rho_{\rm crit}}\right)^{1/3}.
\end{equation}
We note that our subhalo profile has a central cusp, unlike the cored density profile considered in \cite{Williams2023FlashlightsDMSubhalo}.

For a given subhalo with mass $m$ and location $R$, we find the parameters $(m_{200}, c_{200}, r_\mathrm{t})$ in the following way:  we first identify the tidal truncation radius with the instantaneous tidal radius,
\begin{equation}
    r_\mathrm{t}(m, R) = R \left(\frac{m}{M_<(R)}\right)^{\frac13} \left(3 - \frac{\rmd \ln M_<(R)}{\rmd \ln R}\right)^{-\frac13},
\end{equation}
where $M_<(R)$ is the host-halo mass enclosed within radius $R$.  We then determine the subhalo's underlying NFW concentration $c_{200}$ using the phenomenological relation \citep{2001MNRAS.321..559B, 2007ApJ...667..859D, 2015PhRvD..92l3508B, 2017MNRAS.466.4974M}
\begin{equation}
    c_{200}(m, R) = \overline C_{200}(m) \left\{1 + \frac{1}{15} \left[\frac{1.5^2}{0.1^2 + (R/R_{200})^2}\right]^\frac{1}{2}\right\},
\end{equation}
where $R_{200}$ is the host-halo radius and $\overline C_{200}(m)$ is the mean concentration for field halos as described in \citet{2016MNRAS.460.1214L}. Finally, we solve for an underlying NFW mass $m_{200}$ that reproduces the correct total mass $m = \int_0 ^{\infty} 4 \pi r^2 \rmd r\, \rho_\mathrm{SH}(r)$.

The deflection caused by a subhalo with parameters $(m_{200}, c_{200}, r_\mathrm{t})$ located at $\vec x_0$ is given by
\begin{equation}
    \vec \alpha_{\rm SH}(\vec x) = \frac{4\,G\,m_{200}}{c^2\, f(c_{200})\,r_\mathrm{s}} \frac{D_\mathrm{LS}}{D_\mathrm{S}} \frac{\vec r}{|\vec r|} S\left(\frac{D_\mathrm{L}}{r_\mathrm{s}}|\vec r|, \frac{r_\mathrm{t}}{r_\mathrm{s}} \right),
\end{equation}
where $\vec r \equiv \vec x - \vec x_0$ and $S(\xi, \tau)$ is a dimensionless function, whose explicit definition and method of evaluation are detailed in \refapp{subhalo-structure}. In the above, $D_\mathrm{L}$, $D_\mathrm{S}$, and $D_\mathrm{LS}$ are the angular diameter distances to the lens, to the source, and from the lens to the source, respectively. We also provide analytic expressions for the first and second derivatives of the deflection angle in \refapp{subhalo-structure}.

\section{Observables}
\label{sec:observables}

Through the ray equation \refeq{ray-equation}, subhalos perturb the smooth fold caustic and, in the most dramatic cases, break it into a corrugated collection of smaller fold caustics. We shall refer to those as the ``subfolds''. The fact that Earendel does not show as a resolved pair of images suggests that it is nearly exactly on one of the subfolds, if the original smooth macro fold caustic is disrupted. However, the subfolds, on average, have weaker caustic strengths than the smooth fold. This biases the size constraint and hence affects the interpretation of Earendel's astrophysical nature. In this section, we define several important observables of lensing.

\subsection{Size constraint}
\label{sec:size-constraints}
Earendel appears as a single, unresolved source. In the analysis of \citet{Welch2022EarendelNature}, where only the smooth fold is accounted for, one expects that the two lensed images of Earendel straddle the smooth critical curve at $\vec x_{1, 2} = \pm \Delta x\, \vec e_\parallel$. We shall adopt the agnostic assumption that the intrinsic surface-brightness profile of Earendel is a 2-dimensional isotropic Gaussian function with an angular standard deviation $\sigma_\mathrm{S}$. Then each image will be an elongated 2-dimensional Gaussian profile with size $|\mu_\parallel|\sigma_\mathrm{S}$ and $|\mu_\perp| \sigma_\mathrm{S}$ along the degenerate and the perpendicular direction, respectively. Since $|\mu_\parallel| \gg |\mu_\perp|$ near the critical curve, the lensed images are flattened to 1-dimensional Gaussians with an angular standard deviation $|\mu_\parallel|\sigma_\mathrm{S}$. 

As neither the image-pair separation nor the image size is astrometrically resolved, it is required that
\begin{equation}\label{eq:size-constraint-equations}
    \Delta x \lesssim \mathcal R \quad \text{and} \quad |\mu_\parallel| \,\sigma_\mathrm{S} \lesssim \mathcal R,
\end{equation}
where $\mathcal R$ represents the angular resolution ($\simeq 0.055''$ for Hubble Space Telescope as tested by \citet{Welch2022EarendelNature}) of the observation, as quantified by the size of the point-spread function (PSF). Using \refeq{smooth-fold-mu_para-mu_perp}, it follows from \refeq{size-constraint-equations} that the proper size $r_\mathrm{S} \equiv \sigma_\mathrm{S} D_\mathrm{S}$ of Earendel is constrained to be 
\begin{equation} \label{eq:size-constraint}
    r_\mathrm{S} \lesssim \mathcal R ^ 2 D_\mathrm{S} \left| \vec d^ {\vec 0} \cdot \vec e^{\vec 0}_\parallel \right|.
\end{equation}

Beyond this crude analysis, it is non-trivial to determine if an arbitrary set of images will be exactly unresolved (i.e.\ consistent with a point source) under a given PSF. To facilitate better comparison between different methods that we will use to investigate this observable in \refsec{methods}, we shall use the photometric moments for quantification. 

Specifically, the source surface-brightness profile is
\begin{equation}
\label{eq:bs}
    b(\vec s) = \frac{1}{2 \pi\, \sigma_\mathrm{S}^2} \exp\left(-\frac{s^2}{2\,\sigma_\mathrm{S}^2}\right),
\end{equation}
where $\vec s \equiv \vec y - \vec y_\mathrm{S}$ with $s \equiv |\vec s|$ and $\vec y_\mathrm{S}$ is the center of the source.  The surface brightness observed at image position $\vec x$ is then $b(\vec y(\vec x) - \vec y_\mathrm{S})$. 

Along the direction of elongation, we define the photometric moment (a two-by-two matrix)
\begin{equation}\label{eq:photometric-moment-x}
    \mat \Sigma_x \equiv \frac{\int (\vec x - \vec C_x) (\vec x - \vec C_x) b[\vec y(\vec x) - \vec y_\mathrm{S}]\, \rmd^2 x}{\int b[\vec y(\vec x) - \vec y_\mathrm{S}]\, \rmd^2 x},
\end{equation}
where $\vec C_x$ is the photometric center defined as
\begin{equation}\label{eq:photometric-center-x}
    \vec C_x \equiv \frac{\int \vec x\, b[\vec y(\vec x) - \vec y_\mathrm{S}] \, \rmd^2 x}{\int b[\vec y(\vec x) - \vec y_\mathrm{S}]\, \rmd^2 x}.
\end{equation}
The size $\sigma_x$ of a set of images is then defined using the photometric moment $\Sigma_x$ as
\begin{equation}\label{eq:size-x}
    \sigma_x \equiv \left[\varrho(\mat \Sigma_x)\right]^{1/2},
\end{equation}
where $\varrho$ is the spectral-radius operator that returns the maximum of the absolute values of the matrix eigenvalues.

We define that the set of images is unresolved if $\sigma_x < \mathcal R$. Under this definition, the size constraint $\overline \sigma_\mathrm{S}$ is defined such that all source with size $\sigma_\mathrm{S} > \overline \sigma_\mathrm{S}$, for any center position $\vec y_\mathrm{S}$, would have been resolved. Let $\sigma_x(\vec y_\mathrm{S}, \sigma_\mathrm{S})$ be the size defined in \refeq{size-x} as a function of $\vec y_\mathrm{S}$ and $\sigma_\mathrm{S}$. Then $\overline \sigma_\mathrm{S}$ can be expressed as
\begin{equation}\label{eq:size-visible-y-constraint-naive}
    \overline \sigma_\mathrm{S} \equiv \min \{\sigma_\mathrm{S} | \, \forall \vec y_\mathrm{S}, \sigma_x(\vec y_\mathrm{S}, \sigma_\mathrm{S}) \geq \mathcal R\}.
\end{equation}

However, there is one subtlety that compels us to reconsider the definition presented above. There is a half-plane region in the source plane such that, for any $\vec y$, the ray equation, \refeq{ray-equation}, admits no solution $\vec x$ \citep{1992grle.book.....S}.  This means that no ray originating from this region will arrive at the observer (other modestly magnified lensed images may still form, with the corresponding rays passing the lens plane somewhere else on the lensed arc and far away from the particular critical curve we concern here). We term this region the external side of the fold caustic. Because of this, it is in fact possible to have an arbitrarily large source, with nearly all of it hiding on the external side. In this case, only a tiny portion of the source is visible, which can be unresolved. This was noted as a difficult corner case in \citet{Welch2022EarendelNature} but was not further considered due to its small probability. This ambiguity can be removed if we instead constrain the size of the \emph{visible} portion of the source along the degenerate direction $\vec e^{\vec 0}_\parallel$. We define the source-plane photometric moment for the visible portion of the source as
\begin{equation}\label{eq:photometric-moment-visible-y}
    \mat \Sigma_{\mathrm{V},y} \equiv \frac{\int_\mathrm{V} (\vec y - \vec C_{\mathrm{V},y})(\vec y - \vec C_{\mathrm{V},y}) b(\vec y - \vec y_\mathrm{S})\, \rmd^2 y}{\int_\mathrm{V} b(\vec y - \vec y_\mathrm{S})\, \rmd^2 y},
\end{equation}
where the subscripts $\mathrm{V}$ indicate that the integral is performed only over the visible portion of the source. The photometric center of the visible portion of the source is defined as
\begin{equation}\label{eq:photometric-center-visible-y}
    \vec C_{\mathrm{V}, y} \equiv \frac{\int_\mathrm{V} \vec y\, b(\vec y - \vec y_\mathrm{S}) \, \rmd^2 y}{\int_\mathrm{V} b(\vec y - \vec y_\mathrm{S}) \, \rmd^2 y}.
\end{equation}
We now define the 1-dimensional size for the visible portion of the source
\begin{equation}\label{eq:size-visible-y}
    \sigma_\mathrm{V} \equiv \left[\vec e^{\vec 0}_\parallel \cdot \Sigma_{\mathrm{V},y} \cdot \vec e^{\vec 0}_\parallel \right]^{1/2}.
\end{equation}
This is measured along the direction of elongation $\vec e^{\vec 0}_\parallel$ because astrometric constraint is only possible in this direction along which the source is hugely magnified. 

In the rest of the paper, we shall use $\sigma_\mathrm{V}$ in place of $\sigma_\mathrm{S}$. We follow the same procedure in obtaining the constraints $\overline \sigma_\mathrm{V}$ on $\sigma_\mathrm{V}$ as
\begin{equation}\label{eq:size-visible-y-constraint}
    \overline \sigma_\mathrm{V} \equiv \min \{\sigma_\mathrm{V} | \, \forall \vec y_\mathrm{S}, \sigma_x(\vec y_\mathrm{S}, \sigma_\mathrm{V}) \geq \mathcal R\},
\end{equation}
where $\sigma_x=\sigma_x(\vec y_\mathrm{S}, \sigma_\mathrm{V})$ is now a function of $\vec y_\mathrm{S}$ and $\sigma_\mathrm{V}$, not $\sigma_\mathrm{S}$.  According to \refapp{detailed-size-constraints}, the constraint on the size of the visible portion of the source $r_\mathrm{V} \equiv D_\mathrm{S}\,\sigma_\mathrm{V}$ is
\begin{equation}\label{eq:physical-size-constraint}
    r_\mathrm{V} < \overline r_\mathrm{V} \equiv D_\mathrm{S}\, \overline \sigma_\mathrm{V} = \mathcal R ^ 2 D_\mathrm{S} \left| \vec d^ {\vec 0} \cdot \vec e^{\vec 0}_\parallel \right|.
\end{equation}
This result agrees with that of the crude analysis in \refeq{size-constraint}. Nevertheless, we emphasize that \refeq{physical-size-constraint} is the exact size constraint for the visible portion of the source.

Using numerical values from \citet{Welch2022EarendelNature}, we derive
\begin{equation}
     \overline r_\mathrm{V} \simeq 0.48\, \mathrm{pc}\, \left(\frac{\mathcal R}{0.055''}\right)^2 \left(\frac{D_\mathrm{S}}{1182\, \mathrm{Mpc}}\right) \left( \frac{\left| \vec d^ {\vec 0} \cdot \vec e^{\vec 0}_\parallel \right|}{\frac{1}{18''} \cdot \cos \frac{\pi}{3}} \right)
\end{equation}
for the conservative choice $D^{\vec 0} = 1 / (18'')$. The size constraint tightens to $\overline r_\mathrm{V} = 0.08\, \mathrm{pc}$ if $D^{\vec 0} = 1/(113'')$. This is the key argument in \citet{Welch2022EarendelNature} that disfavors the star cluster scenario. For a comparison, one of the most compact star clusters found in our cosmic backyard, the R136 cluster in the Large Magellanic Cloud (LMC), has a half-light radius of $\sim 1.7\,$pc~\citep{Hunter1995R136HSTStarCount}. In the Milky Way and M31, globular clusters all have half-light radii larger than $1\,$pc~\citep{Baumgardt2018MWGCcatalog, Barmby2007M31GCcatalog}.  

We will show that the argument presented in \citet{Welch2022EarendelNature} needs to be revised if the smooth fold caustic is disrupted into subfolds due to perturbations from subhalos. Furthermore, given the possibility that more than two images could form with subhalos, we must extend the definition ``unresolved'' to the following:  we first find all unresolved clusters of lensed images (hereafter image clusters) at an angular resolution $\lesssim \mathcal R$; we then require that Earendel must correspond to the brightest image cluster, while the second brightest image cluster, if in existence, must be at least $r_\mathrm{thres} = 10$ times fainter and hence remains undetected. We have chosen this value for the minimum magnification ratio based on the point source detection limit of the JWST/NIRCam images. We will derive a more reliable size constraint following this requirement.

\subsection{Astrometry of Photometric Center}
\label{sec:astrometric-effect}

Subhalos can significantly perturb the position of the lensed images near a macro critical curve~\citep{Dai2018Abell370}.
Judging from Earendel's location relative to the lensed image pair of a compact source on the same arc, whose separation must be bisected by the \emph{smooth} critical curve (each image is nearly $1''$ from Earendel), Earendel shows almost exactly on top of the \emph{smooth} critical curve. This implies that the typical astrometric departure of the perturbed critical curve from the smooth one at Earendel's location cannot be too large, which may constrain the abundance of subhalos. We shall define an astrometric observable to quantify this effect.

Let $\vec c$ be the observed photometric center of Earendel. For a smooth fold caustic we would expect a photometric center $\vec c'$ that is along the degenerate direction and exactly on top of the smooth critical curve:
\begin{equation}
    (\vec c - \vec c') \cdot \vec e^{\vec 0}_\perp = 0 \quad \text{and} \quad \vec D^{\vec 0} \cdot \vec c' = 0,
\end{equation}
respectively.  Solving for $\vec c'$ gives
\begin{equation} \label{eq:photometric-center-smooth-fold}
    \vec c' = (\vec c \cdot \vec e^{\vec 0}_\perp) \left[\vec e^{\vec 0}_\perp - \left(\frac{\vec D^{\vec 0} \cdot \vec e^{\vec 0}_{\perp}}{\vec D^{\vec 0} \cdot \vec e^{\vec 0}_\parallel} \right) \vec e^{\vec 0}_\parallel \right].
\end{equation}
We then define the difference $\Delta c \equiv (\vec c - \vec c') \cdot \vec e^{\vec 0}_\parallel$ of these two predictions along the degenerate direction $\vec e^{\vec 0}_\parallel$ to be the astrometric shift.  Using \refeq{photometric-center-smooth-fold}, we obtain
\begin{equation}
    \Delta c = \vec c \cdot \left[\vec e^{\vec 0}_{\parallel} + \left(\frac{\vec D^{\vec 0} \cdot \vec e^{\vec 0}_{\perp}}{\vec D^{\vec 0} \cdot \vec e^{\vec 0}_\parallel} \right) \vec e^{\vec 0}_\perp \right].
\end{equation}
We emphasize that this vector is along the degenerate direction, not perpendicular to the smooth critical curve. Thus it is suitable in determining, for instance, how far Earendel is from the midpoint of other lensed image pairs along the arc.

\section{Methods}
\label{sec:methods}

We employ two methods, one semi-analytic and another numerical, to derive the key quantitative results of this work. In the first method, we assume that Earendel is located close enough to a subfold, so that the local lensing map can still be approximated by a simple fold, with the caustic strength describing the subfold rather than the smooth macro fold. We refer to this method as the ``subfold approximation''.

In the second method, we exactly compute all lensed images of a finite-sized source via the inverse ray-shooting method augmented with adaptive mesh refinement. We refer to this method as the ``exact numerical'' method. In what follows, we first introduce both methods, and then describe how the observables defined in \refsec{observables} can be extracted.

Both methods start with shooting rays according to \refeq{ray-equation} on an uniform $N_1 \times N_2$ parallelogram mesh
\begin{equation} \label{eq:ray-shooting-mesh}
    \vec x_{i_1 i_2} \equiv l_{1, i_1} \vec n_1 + l_{2, i_2} \vec n_2.
\end{equation}
Here, $i_1 = 1, 2, \cdots, N_1$ and $i_2 = 1, 2, \cdots, N_2$ are the indices for the mesh basis $\vec n_1$ and $\vec n_2$, respectively. And $l_{1, i_1}$ and $l_{2, i_2}$ are uniform nodes on $[-L_1/2, +L_1/2]$ and $[-L_2/2, +L_2/2]$ (including endpoints), respectively. We choose $\vec n_1$ to be along the degenerate direction, i.e.\ $\vec n_1 = \vec e^{\vec 0}_\parallel$, and $\vec n_2$ to satisfy $\vec n_2 \cdot \vec D^{\vec 0} = 0$ so that it is along the smooth critical curve. This choice mitigates the loss of one of the lensed image pair beyond the ray-shooting domain. The area of the domain $\Delta \Omega$ is hence the area of the parallelogram,
\begin{equation}
    \Delta \Omega \equiv L_1 L_2 \sqrt{1 - (\vec n_1 \cdot \vec n_2)^2}.
\end{equation}
We note that $\Delta \Omega$ is different from the field $\Delta S$ of subhalo sampling used in \refeq{subhalo-distribution-2d}. We ensure that the field $\Delta S$ always encloses the field $\Delta \Omega$. For each ray $\vec x_{i_1 i_2}$ shot, we compute and store the deflection $\vec \alpha$, the first-order derivatives $\partial \alpha_i / \partial x_j$, and the second-order derivatives $\partial^2 \alpha_i / \partial x_j \partial x_k$.

\subsection{Subfold Approximation}
\label{sec:subfold-approximation}

Following ray-shooting, we use a contour-finding algorithm to trace out the critical curve $\vec x(t)$ where $\det A(\vec x)$ vanishes. Here, $t$ is the parameter along the critical curve, and we define the subfold $\vec y(t) \equiv \vec x(t) - \vec \alpha[\vec x(t)]$. At $\vec y(t)$, similar to \refsec{smooth-fold}, we define the subfold parameters $(\kappa^t, \vec d^t, \vec e^t_\parallel, \vec e^t_\perp)$, now with the superscript $t$ emphasizing the location of evaluation. We now assume that the source is very close to the subfold and only one pair of highly magnified images are relevant, with the midpoint $\vec x(t)$ on the critical curve.  For each source, we study the following three properties.

\paragraph{Consistency with Earendel} We compute the photometric size $\sigma_x$ of the source as defined in \refeq{size-x} using the analytic expression \refeq{size-x-explicit} provided in \refapp{detailed-size-constraints}. If $\sigma_x < \mathcal R$, we deem the source as consistent with Earendel, and otherwise not.

\paragraph{Visible Proper Size} We compute the visible proper size $\sigma_\mathrm{V}$ of the source as defined in \refeq{size-visible-y} using the analytic expression \refeq{size-visible-y-explicit} provided in \refapp{detailed-size-constraints}, which is appropriate for the fold caustic lens mapping.

\paragraph{Astrometry of the Photometric Center} Under the subfold approximation, the two images formed only differ in parity. Thus the photometric center will lie exactly on the critical curve, i.e.\ $\vec c = \vec x(t)$, giving
\begin{equation}\label{eq:astrometric-effect-subfold}
    \Delta c = \vec x(t) \cdot \left[\vec e^{\vec 0}_{\parallel} + \left(\frac{\vec D^{\vec 0} \cdot \vec e^{\vec 0}_{\perp}}{\vec D^{\vec 0} \cdot \vec e^{\vec 0}_\parallel} \right) \vec e^{\vec 0}_\perp \right].
\end{equation}

Note that now $\sigma_x$, $\sigma_\mathrm{V}$, and $\Delta c$ are all stochastic variables dependent on the subhalo realization drawn from the population model in \refsec{subhalo-abundance}. In practice, we approximate the ensemble statistics as statistics of the parameter $t$ weighted by the subfold length $|\rmd \vec y / \rmd t|\, \rmd t$.

\subsection{Exact Numerical}
\label{sec:exact-numerical}

The subfold approximation is straightforward but only accounts for one pair of lensed images. This approximation may also be inaccurate if the source's distance to the subfold is greater than the source-plane length scale over which the simple fold is a good approximation. The latter problem becomes particularly severe when lower-mass subhalos induce increasingly smaller and numerous wiggles along the critical curve, rendering the critical curve essentially a fractal.

To improve upon the subhalo approximation, we develop an image-finding algorithm using adaptive mesh refinement, which follows the algorithm described in \citet{2014MNRAS.445.1942M}. We term this algorithm the ``telescoping'' algorithm. When the ray equation \refeq{ray-equation} is given, the algorithm finds all images with magnification $|\mu| \gtrsim \mu_\mathrm{min}$ for a finite-sized source.

Let $b(\vec s)$ be the surface brightness profile of the source, where $\vec s \equiv \vec y - \vec y_\mathrm{S}$ is the displacement from the source center $\vec y_\mathrm{S}$. We start by choosing a circular disk centered at $\vec y_\mathrm{S}$ with a large radius $\Delta y_\mathrm{S}$ that circumscribes the source. We then consider a sequence of concentric disks with decreased radii $\Delta y_n = \Delta y_0 / 3^n$ starting with $\Delta y_0 = \Delta x_0 / \mu_\mathrm{min}^{1/2}$, where $\Delta x_0$ is the spatial resolution of the initial ray-shooting mesh, \refeq{ray-shooting-mesh}.  For the $n$-th step, rays that hit the circular disk with radius $\Delta y_n$ are kept. The parallelogram cells that are occupied by these rays are individually divided into $3 \times 3$ congruent parallelogram subcells, and new rays are shot at the center of each subcell. The new rays, together with the old rays that hit, are then tested against a circular disk with radius $\Delta y_{n+1}$ in the $(n+1)$-th step. This process is iterated until the disk becomes smaller than $\Delta y_\mathrm{S}$, in which case we instead set the disk radius to be $\Delta y_\mathrm{S}$ and collect all rays $\vec x_i$ that hit the disk. We then compute the surface brightness $\{ b_i\}$ associated with rays $\{\vec x_i\}$ with $b_i \equiv b(\vec y_i - \vec y_\mathrm{S})$, where $\vec y_i \equiv \vec x_i - \vec \alpha(\vec x_i)$.

For the 2-dimensional Gaussian soure profile \refeq{bs}, the total unlensed flux is then
\begin{equation}
    I_\mathrm{UL} = 1 - \exp\left(- \frac{\Delta y_S^2}{2\,\sigma_\mathrm{S}^2}\right).
\end{equation}
We choose the bounding circle radius $\Delta y_\mathrm{S} = 3\,\sigma_\mathrm{S}$ so that we retain $99\%$ of the flux. We also set a safe magnification floor $\mu_\mathrm{min} = 1.0$.

Our method is a simplified version of the one described in \citet{2014MNRAS.445.1942M} in that we do not refine the image further after the source size $\Delta y_\mathrm{S}$ is reached.  This is because that the Gaussian surface brightness profile we use is smooth, and sufficient accuracy for the magnification factor is achievable without further refining.

We now determine whether multiple images of a given source are unresolved using the condition defined at the end of \refsec{size-constraints}.  First, we sort groups of adjacent rays into images.  Since the rays $\{\vec x_i\}$ are on a parallelogram mesh, albeit at a finer resolution than that defined in \refeq{ray-shooting-mesh}, we define adjacency using the refined mesh. A depth-first search is carried out to find the $N_J$ adjacent groups of rays $\{\vec x_i | i \in J_j\}$, where $J_j$ with $j = 1, 2, \cdots, N_J$ are sets of ray indices. 

Second, we find the photometric center
\begin{equation}
    \vec c_{J_j} \equiv \langle \vec x_i \rangle_{J_j}
\end{equation}
of image $J_j$ and perform a single-linkage clustering with threshold $\mathcal R$ on the centers $\{\vec c_{J_j}\}$ to find $N_K$ image clusters $\{\{\vec x_i | i \in J_j\} | j \in K_k\}$, where $K_k$ with $k = 1, 2, \cdots, N_K$ are sets of images.  Here we have defined the short-hand notation
\begin{equation}
    \langle f_i \rangle_{J_j} \equiv \frac{\sum_{i \in J_j} b_i\,f_i}{\sum_{i \in J_j} b_i}.
\end{equation}

Finally, the magnification can be computed from
\begin{equation}
    \mu_{K_k} = \frac{1}{I_\mathrm{UL}}\sum_{j \in K_k} \sum_{i \in J_j} b_i\,S_\mathrm{cell}
\end{equation}
of image cluster $K_k$, where $S_\mathrm{cell}$ is the area of the \emph{refined} cell. We sort the image cluster so that $\mu_{K_1} \geq \mu_{K_2} \geq \cdots$ when $N_K \geq 2$.  We also compute
\begin{equation}\label{eq:size-x-numerical}
    l_{K_k} \equiv [\varrho(\Sigma_{K_k})]^{1/2}
\end{equation}
as an estimation for the size of image cluster $K_k$, where $\varrho$ is the spectral radius operator that returns the maximum of the absolute values of the eigenvalues of the photometric-moment matrix
\begin{equation}
    \Sigma_{K_k} \equiv \langle (\vec x_i - \vec c_{K_k}) (\vec x_i - \vec c_{K_k}) \rangle_{K_k}.
\end{equation}
This estimator is designed such that it yields the correct value $\Delta x$ when there are only two rays, with the same surface brightness, separated by $\Delta x$.  Here we have introduced the short-hand notation
\begin{equation}
    \langle f_i \rangle _{K_k} \equiv \frac{\sum_{j \in K_k} \sum_{i \in J_j} b_i\,f_i}{\sum_{j \in K_k} \sum_{i \in J_j} b_i}
\end{equation}
and defined the photometric center
\begin{equation}
    \vec c_{K_k} \equiv \langle \vec x_i \rangle_{K_k}.
\end{equation}

Following these calculations, we discard the source if the photometric center $\vec c_{K_1}$ of the brightest image cluster is in the outer 30\% margin of the ray-shooting mesh. This is to avoid the situation in which some lensed images are clipped off the simulation domain. After this selection, the following three properties are studied for each allowed source.

\paragraph{Consistency with Earendel} The source is deemed consistent with Earendel if either i) $N_K = 1$ and $l_{K_1} < \mathcal R$ or ii) $N_K \geq 2$ but $\mu_{K_1} / \mu_{K_2} > r_\mathrm{thres}$ and $l_{K_1} < \mathcal R$. Here $r_\mathrm{thres} = 10$ is the minimum magnification ratio we choose to rule out a second detectable image cluster elsewhere.

\paragraph{Visible Proper Size} We compute the visible proper size $\sigma_\mathrm{V}$ via the definition \refeq{size-visible-y} by using the rays $\{\vec y_i\}$ as samples to estimate the integral. Since the rays $\{\vec x_i\}$ are uniformly sampled in the image plane, each ray $\vec y_i$ should be down-weighted by the magnification $\mu(\vec x_i)$.

\paragraph{Astrometry of the Photometric Center}  For the unresolved sources, we use the photometric center of the brightest image cluster to compute the astrometric deflection,
\begin{equation}\label{eq:astrometric-effect-numerical}
    \Delta c = \vec c_{K_1} \cdot \left[\vec e^{\vec 0}_{\parallel} + \left(\frac{\vec D^{\vec 0} \cdot \vec e^{\vec 0}_{\perp}}{\vec D^{\vec 0} \cdot \vec e^{\vec 0}_\parallel} \right) \vec e^{\vec 0}_\perp \right].
\end{equation}

\section{Results}
\label{sec:results}

We now apply the methods introduced in \refsec{subfold-approximation} and \refsec{exact-numerical} to study the effect of subhalos on the interpretation of Earendel.

First, we consider three subhalo population scenarios with the minimum infall masses $m_\mathrm{min} = \{10^6, 10^7, 10^8\}\, M_\odot$ but a common maximum infall mass $m_\mathrm{max} = 10^9\, M_\odot$. Subhalos with $m_{\rm min}>10^9\,M_\odot$ have a low surface number density on the lens plane such that possibly Earendel is not affected by any such massive subhalos. Smaller subhalos are expected to be sufficiently numerous that the macro critical curve is perturbed by them everywhere~\citep{Dai2018Abell370, Williams2023FlashlightsDMSubhalo}. Including subhalos with $m_{\rm min}<10^6\,M_\odot$ would be computationally expensive, but they may not imprint significant observable effects in lensing. These subhalos are orders of magnitude less massive than the $m_{\rm min}=10^6$--$10^8\,M_\odot$ ones but not orders of magnitude smaller in size according to the mass-concentration relation predicted by the halo collapse theory. Because of their small convergence and shear, these subhalos are inefficient in creating additional large caustics, and may have limited effects in relaxing the source size constraint for Earendel. Neglecting those very small subhalos in our numerical calculations is therefore a conservative choice. For the three minimum infall masses $m_\mathrm{min} = \{10^6, 10^7, 10^8\}\, M_\odot$ that we consider, the subhalo model produces the mean subhalo convergences $\kappa_\mathrm{SH} = \{11.6, 8.6, 4.8\} \times 10^{-3}$, respectively.

\begin{figure*}
    \centering
    \includegraphics[width = \linewidth]{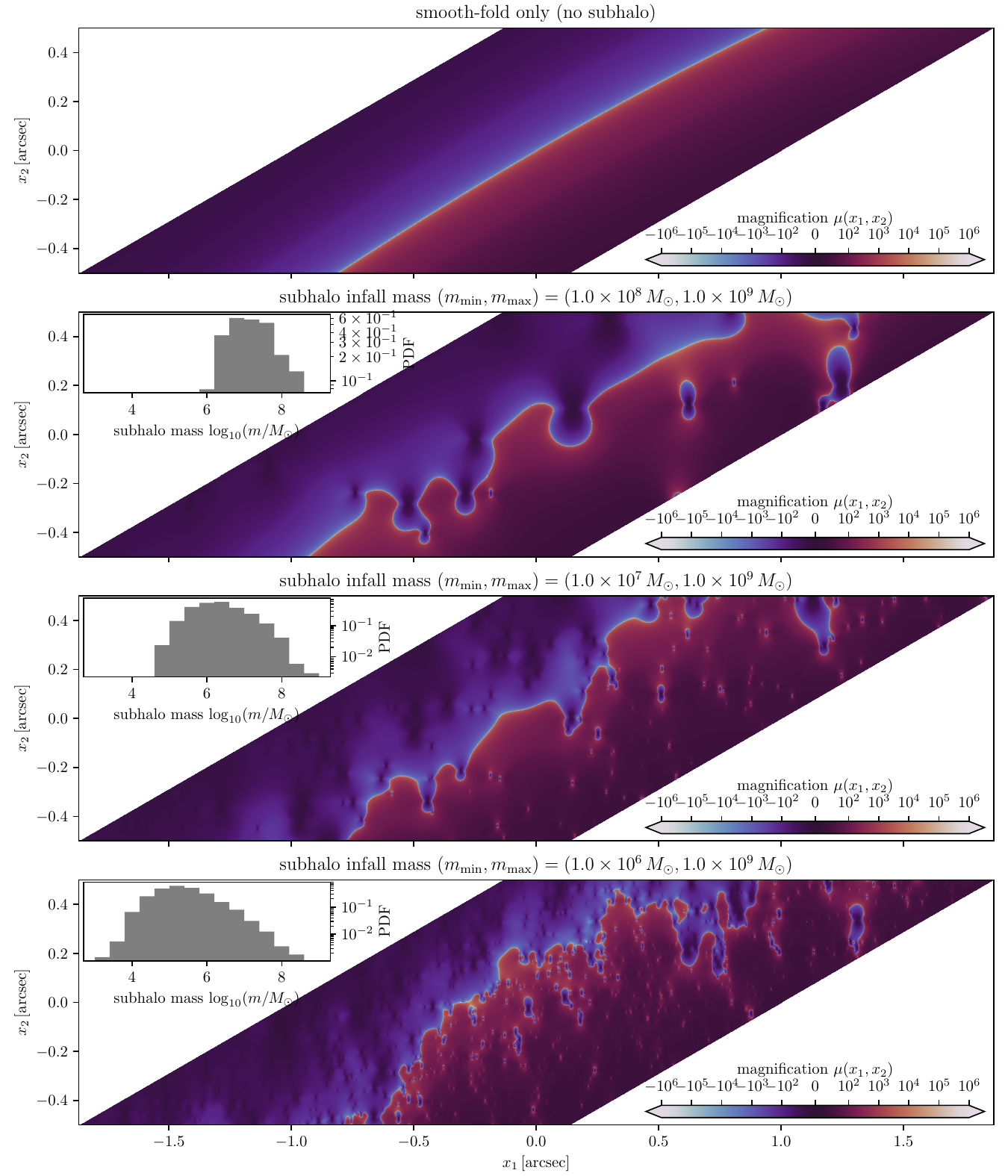}
    \caption{Magnification map for random realizations of subhalos under different subhalo population scenarios in the vicinity of Earendel where the macro critical curve intersects the Sunrise Arc. The top panel shows the case where there is only a single smooth fold, as assumed by \citet{Welch2022EarendelNature}.  The rest, from top to bottom, show the cases where the subhalo \emph{infall} masses are between $m_\mathrm{min} = \{10^8, 10^7, 10^6\}\, M_\odot$ and $m_\mathrm{max} = 10^9\, M_\odot$. Within each panel, the inset histogram shows the (\emph{final}, after tidal disruption) subhalo-mass distribution. Subhalos disturb the shape of the smooth critical curve, and changes the local strength of the fold.}
    \label{fig:field-of-view}
\end{figure*}

The image-plane magnification map for random realizations of these scenarios are shown in \reffig{field-of-view} on a parallelogram domain with the size $(L_1, L_2) = (2'', 2'')$. We sample source center by randomly selecting a point on the caustic and randomly move the point along the degenerate direction $\vec e_\parallel$ with an amount uniformly distributed in $(-10\, \sigma_\mathrm{S}, +10\, \sigma_\mathrm{S})$.  We perform ray-shooting to find the lensed images and determine if it is consistent with the observation of Earendel. We also compute the size for the visible portion of the source $r_\mathrm{V}$ (described in \refsec{size-constraints}) and the astrometric effect $\Delta c$ (described in \refsec{astrometric-effect}). We study four different source sizes $r_\mathrm{S} = \{0.1, 0.3, 1.0, 3.0\}\, \mathrm{pc}$, and set an astrometric resolution $\mathcal R = 0.055''$~\citep{Welch2022EarendelNature} and a magnification-ratio cut $r_\mathrm{thres} = 10$.

\begin{figure*}
    \centering
    \includegraphics[width = \linewidth]{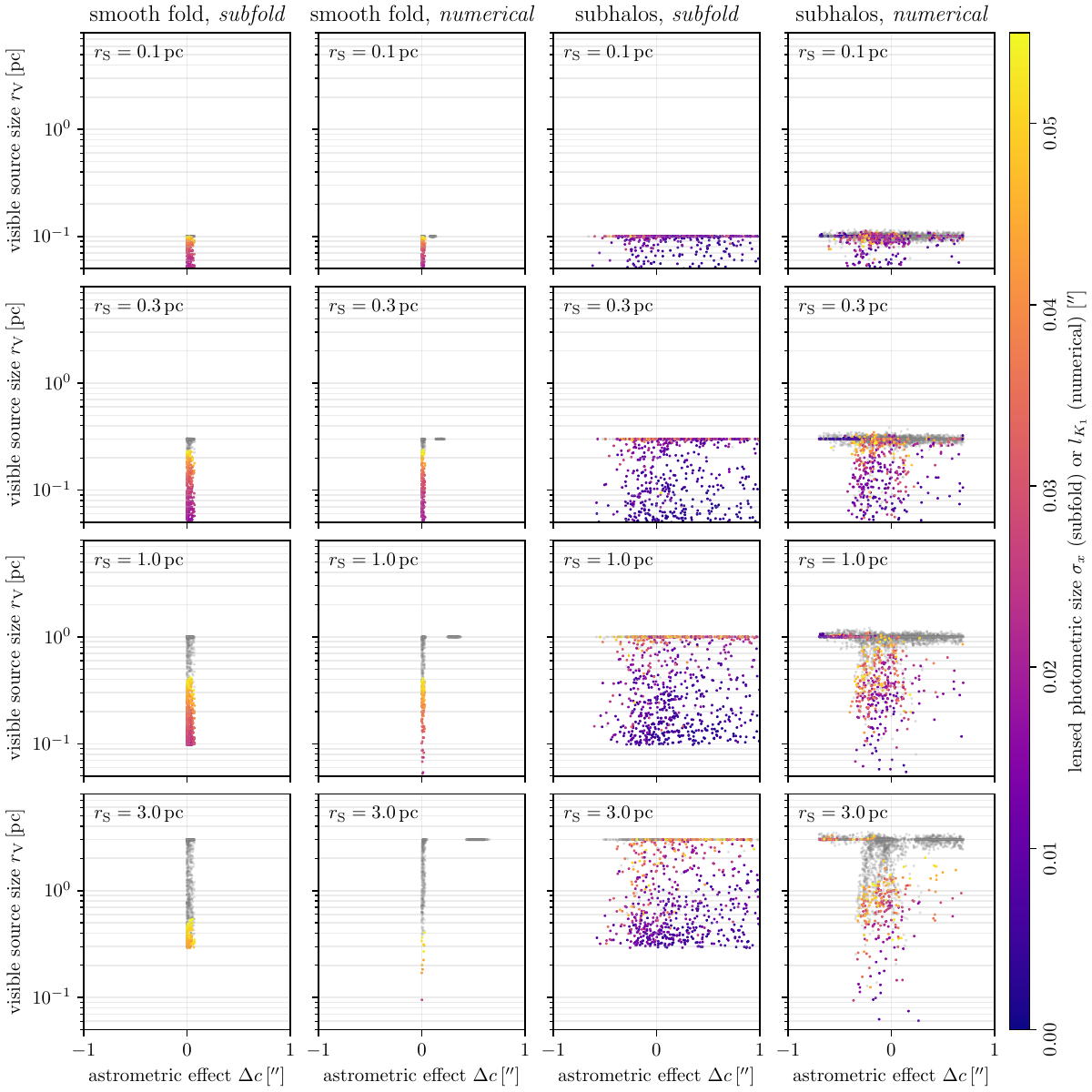}
    \caption{Comparison between the subfold approximation and the exact numerical method on the possible source size of Earendel. The left and the right two columns are for the case with and without the presence of the subhalos, respectively. In the subhalo case, the infall mass is chosen in the range $(m_\mathrm{min}, m_\mathrm{max}) = (10^6, 10^9)\, M_\odot$.  In each case, the first and the second columns are obtained via the subfold approximation and the exact numerical method, respectively. Each panel shows, with color coding, the visible source size $r_\mathrm{V}$ [defined in \refeq{size-visible-y}] versus the astrometric effect $\Delta c$ [defined in \refeq{astrometric-effect-subfold} and \refeq{astrometric-effect-numerical} respectively for the two methods]. We choose sources with sizes $\sigma_\mathrm{S} = \{0.1, 0.3, 1.0, 3.0\}\, \mathrm{pc}$ in each row. Sources compatible with Earendel are color-coded by their lensed photometric size $\sigma_x$ [defined in \refeq{size-x}] or $l_{K_1}$ [defined in \refeq{size-x-numerical}] respectively for the two methods. Sources incompatible with Earendel are colored in gray. In the presence of subhalos, the exact numerical method is needed to obtain accurate results, while the subfold method provides a fair approximation. The presence of subhalos relaxes the constraint on the \emph{visible} size $r_\mathrm{V}$ by a factor of a few to ten.}
    \label{fig:comparison}
\end{figure*}

\reffig{comparison} shows a comparison between the subfold approximation and the exact numerical methods.  In the reference case with only a smooth fold, the two methods yield very close results. When numerous low-mass subhalos are included, with $m_\mathrm{min} = 10^6 \, M_\odot$, the exact numerical method finds less sources to have a single unresolved image. This is because the exact numerical method uses the full lens mapping and a more realistic procedure in determining whether the set of images will appear as a single unresolved image consistent with the observation of Earendel.

\begin{figure*}
    \centering
    \includegraphics[width = \linewidth]{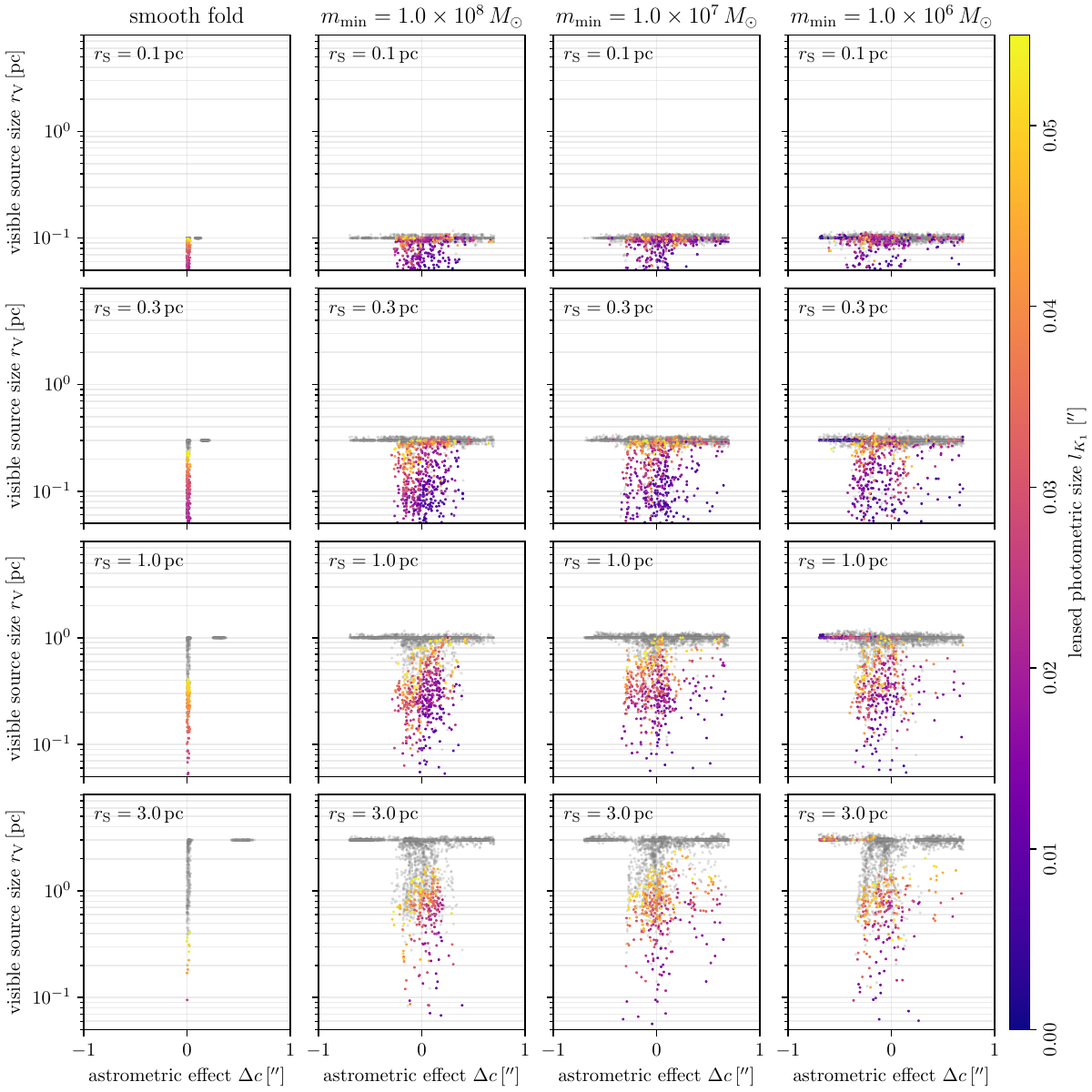}
    \caption{Results derived using the exact numerical method for different subhalo population scenarios. In the second to the fourth columns are the subhalo scenarios with minimum infall masses $m_\mathrm{min} = \{10^6, 10^7, 10^8\}\, M_\odot$ and a maximum infall mass $m_\mathrm{max} = 10^9\, M_\odot$. In the first column, the smooth-fold-only scenario is plotted for reference. All other elements of this figure are the same as those in \reffig{comparison}. The constraint on the visible size $r_\mathrm{V}$ of Earendel is not very sensitive to the minimum cutoff $m_\mathrm{min}$ of the infall mass.}
    \label{fig:results-100}
\end{figure*}

\begin{figure*}
    \centering
    \includegraphics[width = \linewidth]{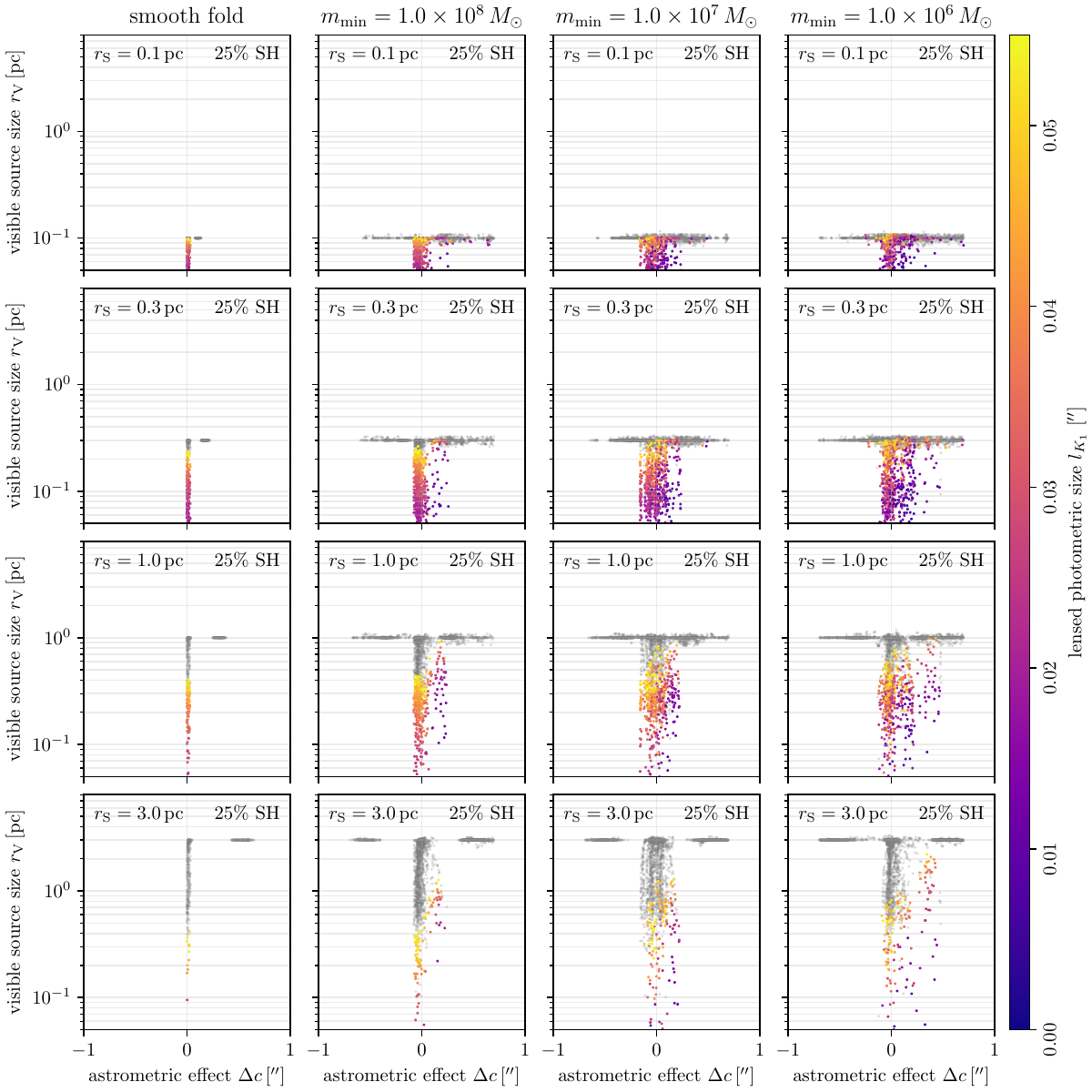}
    \caption{Same as \reffig{results-100}, but with the subhalo abundance reduced to 25\% of the fiducial model. The constraint on the visible size $r_\mathrm{V}$ of Earendel does not significantly tighten with this reduction of the subhalo abundance.}
    \label{fig:results-025}
\end{figure*}

Since the exact numerical method is a more accurate treatment for the lensing effect of subhalos, next we only discuss results derived from the exact numerical method. \reffig{results-100} compare results derived from the exact numerical method for three different values of the minimum subhalo infall masses $m_\mathrm{min} = \{10^6, 10^7, 10^8\}\, M_\odot$. \reffig{results-025} is the same with \reffig{results-100}, but  with the subhalo abundance reduced to $25\%$ of the fiducial value. For these scenarios, the mean subhalo convergence is then $25\%$ that of the fiducial model.

These results show that the constraints to the visible size $r_\mathrm{V}$ of Earendel is relaxed by a factor of a few to ten when the effect of subhalos is accounted for. This conclusion is robust when the infall mass cutoff $m_\mathrm{min}$ or the overall subhalo abundance are varied. In particular, the visible portion of Earendel can be larger than $\sim 1\, \mathrm{pc}$, thus allowing the interpretation of Earendel being a compact star cluster. For example, there are many star clusters in local spiral galaxies with a half-light radius $\sim 1$--$2\,$pc (Figure 9 of \cite{KMBH2019ARAAreview}).

We also found that subhalos can perturb the astrometry of the lone image so that it does not sit exactly atop the smooth macro critical curve. However, we find that this astrometric shift is typically $\lesssim 0.5''$ and has a distribution that peaks at $\Delta c = 0''$. Since it is highly expected from cold dark matter that the population of sub-galactic subhalos exists, we conclude that the source size constraint alone cannot rule out star cluster as the possible underlying nature of Earendel.

\section{Discussion}
\label{sec:discussion}

So far we studied how dark matter subhalos, a prediction of the vanilla CDM model, will relax the constraints to the physical size of Earendel.  However, other particle physics models that produce small-scale dark matter clustering will have similar effects.

A notable model of this kind is ultra-light boson dark matter. With a suitable boson mass $m_a$, it can produce de~Broglie interference patterns on a scale similar to the subhalos considered here.  The de~Broglie wavelength of the dark matter particle is
\begin{equation}
    \lambda _\mathrm{dB} = \frac{h}{mv} = 12\, \mathrm{pc} \left(\frac{10^{-21}\, \mathrm{eV}}{m}\right) \left(\frac{1000\, \mathrm{km}/\mathrm{s}}{v}\right),
\end{equation}
where $h$ is the Planck constant and $v$ is the typical virial velocity of the dark matter particles inside the galaxy cluster halo.  The $10^5\, M_\odot$ subhalo that we consider has a scale radius $r_\mathrm{S} \sim 80\, \mathrm{pc}$, which is several times larger than $\lambda_\mathrm{dB}$ for a boson with mass $m_a \simeq 10^{-21}\, \mathrm{eV}$.  For similar boson masses, the effect of interference on lensing in the vicinity of macro critical curves has been previously investigated \citep{Amruth2023EinsteinRingWaveDM, Diego2024arXiv240608537D, Broadhurst2024arXiv240519422B}.

However, if the boson mass is much larger, the length scale of the interference pattern is small, and a large number of small wave interference structures integrated along the line of sight will reduce both the magnitude and coherence length of the lens surface density fluctuations. Recently, tight constraints on ultralight boson mass $m_a \gtrsim 10^{-19}\,$eV have been derived from the study of stellar dynamics in ultra-faint dwarf galaxies~\citep{Marsh2019FuzzyDMBoundEridanusII, Dalal2022FuzzyDMbounds}.  Since each $\lambda_\mathrm{dB}$ patch has order-unity density fluctuation, the convergence fluctuation $\delta\kappa$ over distance $L$ is roughly $\delta \kappa \sim (\lambda_\mathrm{dB} / L)^{1/2}$, giving
\begin{equation}
    \delta \kappa \sim 10^{-3}\,\left(\frac{m}{10^{-19}\,{\rm eV}}\right)^{-\frac12}\,\left(\frac{v}{10^3\,{\rm km}\,{\rm s}^{-1}}\right)^{-\frac12}\,\left(\frac{L}{100\,{\rm kpc}}\right)^{-\frac12}.
\end{equation}
for a line-of-sight projection across $L \sim 100\,$kpc of the dense part of the galaxy cluster dark matter halo. If the boson mass bounds quoted above are valid, ultra-light dark matter may be relevant on tiny angular scales analogous to and subdominant than intracluster microlensing, but is not expected to generate the same larger-scale, persistent effect from CDM subhalos as we have studied here.

\section{Conclusion}
\label{sec:concl}

In this paper, we studied the effect of dark matter subhalos on interpreting highly magnified sources on lensing caustics. This is a sub-galactic population with subhalo masses in the range $10^6$--$10^9\,M_\odot$ that are expected to arise from hierarchical structure formation in the cold dark matter model but are thought to lack the stellar component. We consider the particular example of the highly magnified source Earendel, discovered in a lensed galaxy at $z_s = 6.2$ as a case study, as it has been suspected to be an extremely magnified single or multiple star system but a crucial constraint on the underlying source size was previously derived based on lensing analysis neglecting the presence of sub-galactic subhalos. 

We have recreated the smooth lens model in the vicinity of Earendel as reported in \citet{Welch2022EarendelNature} and have reproduced the previously reported tight size constraint. We have clarified that the size constraint is actually on the visible portion of the source when the source overlaps the caustic, which is not an improbable corner case for large sources. We then embedded in the smooth lens model randomly generated subhalos according to calibrated semi-analytic models of the subhalo internal structure and population. Such lensing situations with random realizations of subhalos are then numerically examined under the semi-analytical subfold approximation as well as with the exact numerical ray-shooting method.

We find that the presence of subhalos relaxes the constraint on the size of the visible portion of Earendel by a factor of a few to ten compared to that derived for the smooth lens model. This aligns with the intuition that small-scale perturber lenses weaken and disrupt lensing caustics despite their small contributions to convergence and shear. Our results imply that the physical size of Earendel may be larger than $\sim 1\, \mathrm{pc}$ without contradicting observation, making it a viable possibility that Earendel is a compact star cluster with a large number of stars. We point out that subhalos can also astrometrically perturb the lensed image of Earendel so that it may not necessarily sit exactly atop the smooth critical curve.  However, we find that the deviation is typically smaller than $\sim 0.5''$ and having nearly zero astrometric departure is a probable realization. Our findings are robust to moderate changes in the assumed subhalo population, such as the abundance of subhalos or the lower cutoff on the subhalo mass.

Based on JWST NIRCam photometry, \cite{Welch2022EarendelJWSTimaging} found that if Earendel is visually dominated by a single star a cool supergiant with $T_{\rm eff}=15000\,$K would be the best fit. However, the authors noted that the single star SED fit had significant discrepancy with data and explored a scenario that Earendel is a binary consisting of one cool supergiant and one O star. The challenge with the binary interpretation is that the hot O star is required to have a higher bolometric luminosity, which is at odds with the expectation that the more massive binary component evolves off the main sequence faster (see \cite{Nabizadeh2024arXiv240612607N} for possible solutions to this problem). This difficulty, together with the non-detection of microlensing-induced time variability, may imply instead that Earendel's light is dominated by many stars, but we note that SED fitting to star cluster models has not been presented in the literature and would be a useful investigation toward unveiling Earendel's true nature.

The idea and methodology of this work can be generalized to a few other recently reported highly-magnified lone images atop critical curves in other lensed galaxies~\citep{Diego2022Godzilla, Furtak2023arXiv230800042F}. We expect that such analysis will impact the interpretation on the nature of these sources as source size constraints may have to be revised. Analyses of more similar highly magnified sources in the future should account for the possible effect of dark matter subhalos when applying the smooth lens model to deducing de-lensed properties. The time-varying microlensing effect of the intracluster stars, which is inevitable and can have an observable effect on both clusters of stars or individual stars, is a complementary approach to constraining de-lensed source properties.

\section*{acknowledgments}
The authors would like to thank Shude Mao, Christopher McKee, Massimo Pascale and Wenrui Xu for useful discussions. L.D. acknowledges research grant support from the Alfred P. Sloan Foundation (Award Number FG-2021-16495), and support of Frank and Karen Dabby STEM Fund in the Society of Hellman Fellows. L.D. is also supported by the Office of Science, Office of High Energy Physics of the U.S. Department of Energy under Award Number DE-SC-0025293.

\appendix

\section{Details of the Lens Models}
\label{app:detailed-lens-models}

We are in the regime of a single lens plane where all deflections of light rays happens in a region whose thickness is negligible compared to the angular diameter distances $D_{S}$, $D_{L}$, and $D_{LS}$ involved. In this case, the total deflection $\vec \alpha(\vec x)$ in the ray equation $\vec y(\vec x) = \vec x - \vec \alpha(\vec x)$ is simply the sum of deflections caused by individual lenses,
\begin{equation}
    \vec \alpha (\vec x) = \vec \alpha_{\rm FD}(\vec x) + \sum_{n=1}^{N} \vec \alpha^{(n)}_{\rm SH}(\vec x) + \vec{\alpha}_{\rm UD}(\vec x),
\end{equation}
where $\vec \alpha_{\rm FD}(\vec x)$, $\vec \alpha_{\rm SH}^{(n)}(\vec x)$, and $\vec{\alpha}_{\rm UD}(\vec x)$ are the deflections from the smooth fold catastrophe, from the $n$-th subhalo, and from the compensating uniform disk, respectively. To evaluate the Jacobian ${\partial y_i}/{\partial x_j}(\vec x) = \delta_{ij} - {\partial \alpha_i}/{\partial x_j}(\vec x)$, we need the first-order derivatives of the individual deflections (which is the second-order derivatives of the Fermat potential),
\begin{equation}
    \frac{\partial \alpha_i}{\partial x_j} (\vec x) = \frac{\partial \alpha_{i, \rm FD}}{\partial x_j}(\vec x) + \sum_{n=1}^{N} \frac{\partial \alpha^{(n)}_{i, \rm SH}}{\partial x_j}(\vec x) + \frac{\partial \alpha_{i, \rm UD}}{\partial x_j} (\vec x),
\end{equation}
and the second-order derivatives as well (which is the third-order derivatives of the Fermat potential),
\begin{equation}
    \frac{\partial^2 \alpha_i}{\partial x_j \partial x_k} (\vec x) = \frac{\partial^2 \alpha_{i, \rm FD}}{\partial x_j \partial x_k}(\vec x) + \sum_{n=1}^{N} \frac{\partial^2 \alpha^{(n)}_{i, \rm SH}}{\partial x_j \partial x_k}(\vec x) + \frac{\partial^2 \alpha_{i, \rm UD}}{\partial x_j \partial x_k} (\vec x),
\end{equation}
In what follows, we collect the formulas we use for each lens model.

\subsection{Fold Catastrophe}
\label{app:fold-catastrophe}

We start by expanding the lensing potential $\psi(\vec x)$ to cubic order at the image-plane origin $\vec x = \vec 0$, which is on the critical curve:
\begin{equation}
    \psi(\vec x) = \psi^{\vec 0}  + \psi^{\vec 0}_i\,x_i + \frac{1}{2!}\,\psi^{\vec 0}_{ij} \,x_i\,x_j + \frac{1}{3!}\,\psi^{\vec 0}_{ijk}\,x_i\,x_j\,x_k + \mathcal O(|\vec x|^4).
\end{equation}
Here, we have adopted the Einstein summation convention and the short-hand notation for derivatives $\psi_{ij\cdots} \equiv \partial \psi / (\partial x_i \partial x_j \cdots)$.  The superscript $\vec 0$ indicates that the quantities are evaluated at the origin $\vec x = \vec 0$. The resulting deflection is
\begin{equation}
    \alpha_{i, \rm FD} (\vec x) = \psi_i(\vec x) = \psi^{\vec 0}_i + \psi^{\vec 0}_{ij}\,x_j + \frac{1}{2!}\,\psi^{\vec 0}_{ijk}\,x_j\,x_k + \mathcal O(|\vec x|^3).
\end{equation}
Without loss of generality, we require that the lens-plane origin $\vec x = \vec 0$ is mapped to the source-plane origin $\vec y = \vec 0$, i.e.\ $\vec \alpha(\vec 0) = \vec 0$, for which we set $\psi ^{\vec 0}_i = 0$. We define the Fermat potential $\phi(\vec x, \vec y) \equiv (\vec x - \vec y)^2 /2 - \psi(\vec x)$, and switch to using derivatives of the Fermat potential $\phi^{\vec 0}_{ij} = \delta_{ij} - \psi^{\vec 0}_{ij}$ and $\phi^{\vec 0}_{ijk} = - \psi^{\vec 0}_{ijk}$, instead of those of the lensing potential, to characterize the fold.  Doing so, we have
\begin{equation}
    \alpha_{i, \rm FD} (\vec x) = x_i - \phi^{\vec 0}_{ij}\,x_j - \frac{1}{2!}\,\phi^{\vec 0}_{ijk}\,x_j\,x_k + \mathcal O(|\vec x|^3),
\end{equation}
The requirement that the origin $\vec x = \vec 0$ is on the fold implies that $\det A(\vec 0) = 0$, thus $\phi^{\vec 0}_{11}\,\phi^{\vec 0}_{22} - (\phi^{\vec 0}_{12})^2 = 0$. The Jacobian matrix is
\begin{equation}
    \frac{\partial \alpha_{i, \rm FD}}{\partial x_j}(\vec x) = \delta_{ij} - \phi^{\vec 0}_{ij} - \phi^{\vec 0}_{ijk}\,x_k + \mathcal O(|\vec x|^2).
\end{equation}
The second derivative of the deflection is
\begin{equation}
    \frac{\partial^2 \alpha_{i, \rm FD}}{\partial x_j \partial x_k} (\vec x) = - \phi^{\vec 0}_{ijk} + \mathcal O(|\vec x|).
\end{equation}

A single fold has the lens map
\begin{equation}
    y_i(\vec x) = x_i - \alpha_{i, \rm FD}(\vec x) = \phi^{\vec 0}_{ij}\,x_j + \frac{1}{2!}\,\phi^{\vec 0}_{ijk}\,x_j\,x_k + \mathcal O(|\vec x|^3).
\end{equation}
The Jacobian matrix is
\begin{equation}
    A_{ij}(\vec x) \equiv \frac{\partial y_i}{\partial x_j}(\vec x) = \phi^{\vec 0}_{ij} + \phi^{\vec 0}_{ijk}\,x_k + \mathcal O(|\vec x|^2) = \begin{pmatrix}
    \phi^{\vec 0}_{11} + \phi^{\vec 0}_{111}\,x_1 + \phi^{\vec 0}_{112}\,x_2 & \phi^{\vec 0}_{12} + \phi^{\vec 0}_{121}\,x_1 + \phi^{\vec 0}_{122}\,x_2 \\
    \phi^{\vec 0}_{21} + \phi^{\vec 0}_{211}\,x_1 + \phi^{\vec 0}_{212}\,x_2 & \phi^{\vec 0}_{12} + \phi^{\vec 0}_{221}\,x_1 + \phi^{\vec 0}_{222}\,x_2
    \end{pmatrix} + \mathcal O(|\vec x|^2),
\end{equation}
with the determinant (accurate to the leading order in $\vec x$)
\begin{equation}
    \det A(\vec x) = \vec x \cdot \vec D^{\vec 0} + \mathcal O(|\vec x|^2), \text{\quad where \quad} D_i^{\vec 0} = \phi^{\vec 0}_{22}\,\phi^{\vec 0}_{11i} + \phi^{\vec 0}_{11}\,\phi^{\vec 0}_{22i} - 2\,\phi^{\vec 0}_{12}\,\phi^{\vec 0}_{12i}.
\end{equation}
A fold is therefore characterized by the independent components of $\phi^{\vec 0}_{ij}$ (i.e.\ $\phi^{\vec 0}_{11}$, $\phi^{\vec 0}_{12}$, and $\phi^{\vec 0}_{22}$) and $\phi^{\vec 0}_{ijk}$ (i.e.\ $\phi^{\vec 0}_{111}$, $\phi^{\vec 0}_{112}$, $\phi^{\vec 0}_{122}$, $\phi^{\vec 0}_{222}$), a total of $7$ numbers, minus the singularity constraint $\phi^{\vec 0}_{11}\,\phi^{\vec 0}_{22} - (\phi^{\vec 0}_{12})^2 = 0$. The magnification $\mu(\vec x) \equiv 1/\det A(\vec x)$ is
\begin{equation}
    \mu(\vec x) = \frac{1}{\vec x\cdot \vec D^{\vec 0}} + \mathcal O\left(\frac{1}{|\vec x|^2}\right).
\end{equation}
A fold is defined by a few parameters that depend only on the local property of the lens \citep{SchneiderEhlersFalco1992textbook, Wagner:2019orn, Lin:2022mrn, Wagner:2022ifl}.

\subsection{Abundance of the Subhalos}
\label{app:subhalo-abundance}

$N$-body simulations suggest that the mass and spatial distributions of subhalos follow simple universal laws independent of the host halo mass. Here, we follow the semi-analytical model developed in \citet{Han:2015pua}, which is based on three simple assumptions about subhalo formation. First, the infall mass $m_\mathrm{acc}$ of a subhalo, before any tidal stripping, follows a power law distribution, $\rmd N (m_\mathrm{acc})/ \rmd \ln m_\mathrm{acc} \propto m_\mathrm{acc}^{-\alpha}$.  Second, the spatial distribution of subhalos at a given infall mass $m_\mathrm{acc}$ follows the density profile $\rho(R)$ of the host halo, $\rmd N(R | m_\mathrm{acc}) / \rmd^3 R \propto \rho(R)$, where $R$ is the host-halo-centric distance and $\rmd^3 R = 4 \pi\,R^2\,\rmd R$ is the volume element. Third, tidal forces gradually strip a subhalo of infall mass $m_\mathrm{acc}$ either 1) completely apart with a probability $(1-f_\mathrm{s})$, or 2) to a reduced final mass $m \propto R^\beta\,m_\mathrm{acc}$ (with a normal scatter $\sigma$ in $\ln m$) with probability $f_\mathrm{s}$. These three assumptions are individually tested in \citet{Han:2015pua} for subhalos with mass $m \gtrsim 10^8 M_\odot$. In this work, we assume that these results can be extrapolated to lower subhalo masses .

Together, the joint distribution of subhalo mass and position is [Eq.~(10) of \citet{Han:2015pua}; but a minus sign is missing in front of the power-law index $\alpha$ in that equation.]
\begin{equation}
    \frac{\rmd N(m, R)}{\rmd \ln m\, \rmd^3 R} = A_\mathrm{acc}\,f_\mathrm{s}\,\frac{\rho(R)}{m_0}\, \left[\frac{m}{m_0\,\bar{\mu}(R)}\right]^{-\alpha}\, \frac{e^{\sigma^2\,\alpha^2 / 2}}{2}\,\left\{\erf\left[\frac{\ln\frac{\bar \mu(R)}{\mu_\mathrm{min}(m)} + \alpha\,\sigma^2}{\sqrt{2}\,\sigma}\right] - \erf\left[\frac{\ln \frac{\bar \mu(R)}{\mu_\mathrm{max}(m)} + \alpha\,\sigma^2}{\sqrt{2}\,\sigma}\right]\right\},
\end{equation}
where $A_\mathrm{acc}$ is a dimensionless normalization constant and $m_0 \equiv 10^{10} M_\odot / h$ is the mass scale of normalization.  Here $\mu \equiv m / m_\mathrm{acc}$ is the fractional mass that remains in the subhalo after tidal stripping, so that the third assumption takes the form $\bar \mu(R) \equiv \mu_\star\,(R / R_{200})^\beta$ with a normalization constant $\mu_\star$. Similarly, $\mu_\mathrm{min}(m) \equiv m / m_\mathrm{max}$ and $\mu_\mathrm{max}(m) \equiv m / m_\mathrm{min}$, where $m_\mathrm{min}$ and $m_\mathrm{max}$ are the minimum and maximum cutoff in the power-law distribution of the infall mass $m_\mathrm{acc}$, respectively. In this work, for a galaxy-cluster-sized host halo, we adopt the fiducial values $A_\mathrm{acc} = 0.08$, $\alpha = 0.9$, $f_\mathrm{s} = 0.56$, $\mu_\star = 0.34$, $\beta = 1.0$, and $\sigma = 1.1$~\citep{Han:2015pua}. We choose to fix $m_\mathrm{max} = 10^{9}\,M_\odot$, and explore the effect of varying $m_\mathrm{min}$.

\subsection{Structure of the Subhalos}
\label{app:subhalo-structure}

We adopt the smoothly truncated NFW (tNFW) density profile for subhalos:
\begin{equation}
    \rho_{\rm SH}(r) = \frac{m_{200}}{4\pi\,r_s^3\,f(c_{200})}\,\frac{1}{(r/r_s)(1+r/r_s)^2}\,\frac{1}{1+r^2/r_t^2},
\end{equation}
parameterized by the mass $m_{200}$, the concentration parameter $c_{200}$ of the underlying untruncated NFW halo, and the truncation radius $r_t$. Here, $f(x) \equiv \ln (1+x) - x/(1+x)$, and the scale radius $r_s$ of the untruncated NFW halo is given in terms of $m_{200}$ and $c_{200}$ as
\begin{equation}
    r_s = \frac{1}{c_{200}}\left(\frac{3}{4\pi} \frac{m_{200}}{200\rho_{\rm crit}}\right)^{1/3}.
\end{equation}
Note that the total mass $m$ of the halo differs from $m_{200}$. It is given by
\begin{equation}
m \equiv \int_{0}^{+\infty} 4\pi\,r^2\,\rmd r\, \rho_{\rm SH}(r) = \frac{m_{200}}{f(c_{200})} \frac{\tau^2}{(\tau^2+1)^2} \left[(\tau^2-1)\ln \tau + (\pi - \tau)\tau -1\right],
\end{equation}
where $\tau\equiv r_t / r_s$. A subhalo centered at $\vec x = \vec x_0$ causes a ray deflection
\begin{equation}
    \vec \alpha_{\rm SH}(\vec x) = \frac{4\,G\,m_{200}}{c^2\,f(c_{200})\,r_s} \frac{D_{LS}}{D_S} \frac{\vec r}{|\vec r|} S\left(\frac{D_L}{r_s}|\vec r|, \frac{r_t}{r_s} \right),
\end{equation}
where we have defined $\vec r \equiv \vec x - \vec x_0$ and the dimensionless function
\begin{equation}
    S(\xi, \tau) \equiv \int_{0}^{+\infty} \frac{\xi \,\rmd\eta}{(\eta^2+\xi^2)^{3/2}} H_1\left(\sqrt{\eta^2+\xi^2}, \tau\right)
\end{equation}
with a closed-form expression for $H_1(t, \tau)$
\begin{equation}
    H_1(t, \tau) \equiv \frac{\tau^2}{2\,(1+\tau^2)^2} \left[-\frac{2\,t}{1+t}(1+\tau^2) + 4\tau \arctan \frac{t}{\tau} + 2\,(\tau^2-1)\,\ln[(1+t)\tau] - (\tau^2-1)\,\ln(t^2+\tau^2)\right].
\end{equation}
The Jacobian matrix is
\begin{equation}
    \frac{\partial \alpha_{i, \rm SH}}{\partial x_j}(\vec x) = \frac{4\,G\,m_{200}}{c^2\,f(c_{200})\,r_s} \frac{D_{LS}}{D_S} \left[(\delta_{ij} - \hat r_i \hat r_j)\,\frac{1}{r}\,S\left(\frac{D_L}{r_s}r, \frac{r_t}{r_s} \right) + \hat r_i\,\hat r_j\,\frac{D_L}{r_s} S_1\left(\frac{D_L}{r_s}r, \frac{r_t}{r_s} \right)\right],
\end{equation}
where we define $S_1(\xi, \tau) \equiv \partial S(\xi, \tau) / \partial \xi$, $r\equiv |\vec r|$, and $\hat r \equiv \vec r / r$.  The second-order derivative of the deflection is 
\begin{multline}
    \frac{\partial \alpha_{i, \rm SH}}{\partial x_j \partial x_k} (\vec x) = 
    \frac{4\,G\,m_{200}}{c^2\,f(c_{200})\,r_s} \frac{D_{LS}}{D_S} 
    \left\{ 
    [3\,\hat r_i\,\hat r_j\,\hat r_k - (\delta_{ij}\,\hat r_{k} + \delta_{jk}\,\hat r_{i} + \delta_{ki}\,\hat r_{j})]\,\frac{1}{r^2}\,S \left(\frac{D_L}{r_s}r, \frac{r_t}{r_s} \right) \right. \\ \left.
    + [(\delta_{ij}\,\hat r_{k} + \delta_{jk}\,\hat r_{i} + \delta_{ki}\,\hat r_{j}) - 3\,\hat r_i\,\hat r_j\,\hat r_k]\,\frac{1}{r}\,\frac{D_L}{r_s}\,S_1 \left(\frac{D_L}{r_s}r, \frac{r_t}{r_s} \right)
    + \hat r_i\,\hat r_j\,\hat r_k\,\left(\frac{D_L}{r_s}\right)^2\,S_2 \left(\frac{D_L}{r_s}r, \frac{r_t}{r_s} \right)
    \right\},
\end{multline}
where we define $S_2(\xi, \tau) \equiv \partial^2 S(\xi, \tau) / \partial \xi^2$.

\begin{figure}
    \centering
    \includegraphics[width = \linewidth]{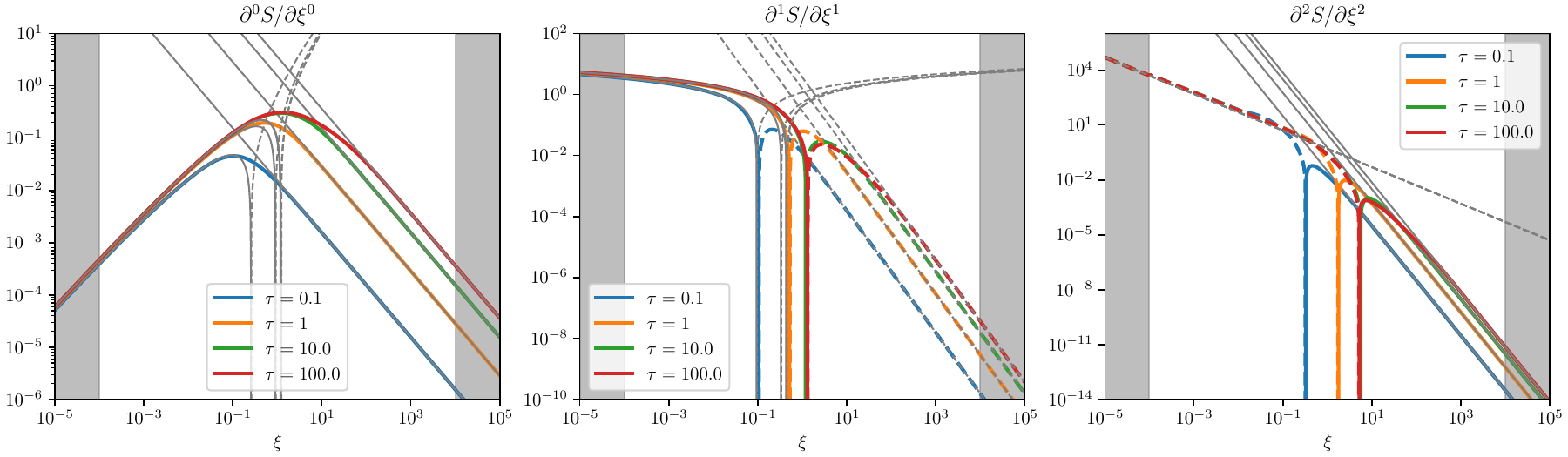}
    \caption{Functions $S(\xi, \tau)$, $S_1(\xi, \tau) \equiv \partial S(\xi, \tau) / \partial \xi$, and $S_2(\xi, \tau) \equiv \partial^2 S(\xi, \tau) / \partial \xi^2$.  The colored lines are for different $\tau$ values and the gray lines are for the corresponding asymptotic formulas for both $\xi\rightarrow 0$ and $\xi\rightarrow +\infty$. The shaded regions in $\xi$ show where the appropriate asymptotic formulas are used. In between the gray regions, uniform rectangular bivariate spline interpolation in logarithmic space is used. Solid (dashed) lines indicate positive (negative) values.}
    \label{fig:S-function-evaluation}
\end{figure}

Although we found analytical expressions for $S(\xi, \tau)$, $S_1(\xi, \tau)$, and $S_2(\xi, \tau)$, they are too lengthy to be computationally efficient. Hence, we proceed with a hybrid scheme of using interpolation and asymptotic formulas. For $(\xi, \tau) \in [10^{-4},\,10^{+4}] \times [10^{-2},\,10^{+3}]$, we tabulate $S(\xi, \tau)$ on a uniform rectangular grid and use bivariate spline interpolation, all in logarithmic space. For values of $\xi$ outside of this range, we directly evaluate the appropriate asymptotic formulas for $\xi \to 0$ or $\xi \to +\infty$. In summary, we have
\begin{equation}
    S(\xi, \tau) = \begin{dcases}
        \left\{\frac{1}{2} \ln \frac{2}{\xi} - \frac{1}{4(1+\tau^2)^2} [\tau^4 + 2(\tau^2 - 1) \ln \tau + 2\pi \tau - 1] \right\} \xi, & \qquad {\rm for}\,\, \xi < 10^{-4}, \\
        \frac{\tau ^2}{\left(\tau ^2+1\right)^2} \left[\left(\tau ^2-1\right) \ln \tau +(\pi -\tau ) \tau -1\right] \xi^{-1}, & \qquad{\rm for}\,\,  \xi > 10^{+4}, \\
        \text{bivariate spline interpolation of $S(\xi, \tau)$} & \qquad \text{otherwise}. 
    \end{dcases}
\end{equation}
For the derivatives $S_1(\xi, \tau)$ and $S_2(\xi, \tau)$, we use the spline derivatives for $\xi \in [10^{-4}, 10^{+4}]$ and the derivatives of the appropriate asymptotic forms outside that range. The accuracy of this method is demonstrated in \reffig{S-function-evaluation}.

\subsection{Uniform Circular Disk Lens}

For a uniform circular disk with angular radius $a$ and dimensionless surface density $\kappa_0$ located at the origin $\vec x = \vec 0$, the ray deflection is
\begin{equation}
    \vec \alpha_\mathrm{UD}(\vec x) = \frac{\vec x}{x} \times \begin{dcases}
        \kappa_0\,x &  (x \leq a) \\
        \kappa_0\,a^2 / x &  (x > a)
    \end{dcases}.
\end{equation}
The first-order and second-order derivatives of the deflection are
\begin{equation}
    \frac{\partial \alpha_{i, \mathrm{UD}}}{\partial x_j}(\vec x) = \begin{dcases}
        \kappa_0\,\delta_{ij} &(x \leq a) \\
        \kappa_0 a^2 \frac{1}{x^2} \left(\delta_{ij} - 2 \frac{x_i x_j}{x^2} \right) & (x > a)
    \end{dcases}
\end{equation}
and
\begin{equation}
    \frac{\partial \alpha_{i, \mathrm{UD}}}{\partial^2 x_j \partial x_k}(\vec x) = \begin{dcases}
        0 &(x \leq a) \\
        \kappa_0 a^2 \left[\frac{8}{x^6} x_i x_j x_k- \frac{2}{x^4} \left(\delta_{ij} x_k + \delta_{jk} x_i + \delta_{ki} x_j\right)\right] & (x > a)
    \end{dcases}
\end{equation}
respectively.

\section{Detailed Size Constraints}
\label{app:detailed-size-constraints}

In this Appendix, we derive in detail the size constraint given in \refsec{size-constraints}.  First, we explicitly compute the photometric center $\vec C_x$, the photometric moment $\Sigma_x$, and the associated image size $\sigma_x$ [defined in \refeq{photometric-center-x}, \refeq{photometric-moment-x}, and \refeq{size-x}, respectively]; then, we explicitly compute the unlensed photometric center $\vec C_{\mathrm{V},y}$, the unlensed photometric moment $\Sigma_{\mathrm{V}, y}$, and the associated unlensed size $\sigma_\mathrm{V}$, all for the visible portion of the source [defined in \refeq{photometric-center-visible-y}, \refeq{photometric-moment-visible-y}, and \refeq{size-visible-y}, respectively]; finally, we derive the size constraint.

Without loss of generality, we choose a coordinate system with basis $(\vec e^{\vec 0}_\parallel, \vec e^{\vec 0}_\perp)$.  In this coordinate, the lensing Jacobian has a simple form,
\begin{equation}
    A(\vec x) = \begin{bmatrix}
        \vec d^{\vec 0} \cdot \vec x & 0 \\
        0 & 2\,(1-\kappa^{\vec 0})
    \end{bmatrix}.
\end{equation}
This indicates that the images will be extremely stretched in the $\vec e_\parallel$ direction, which implies that a 2-dimensional Gaussian source will be stretched to effectively a 1-dimensional Gaussian.  We now focus on the $\vec e_\parallel$ direction.  The lens mapping along the $\vec e^{\vec 0}_\parallel$ direction is obtained by the line integral
\begin{equation}
    \vec y(\lambda) - \vec y(0) = \int _{0}^{\lambda } A[\vec x(\lambda')]\frac{\rmd \vec x(\lambda')}{\rmd \lambda'}\, \rmd \lambda'
\end{equation}
along the image-plane trajectory $\vec x(\lambda ) \equiv \lambda \vec e^{\vec 0}_\parallel$.  With the convention that $\vec y(0) = \vec 0$, we have the 1-dimensional lens mapping along the $\vec e_\parallel$ direction,
\begin{equation}\label{eq:1d-lens-mapping}
    y_\parallel = \frac{1}{2} d^{\vec 0}_\parallel x_\parallel^2,
\end{equation}
where $d^{\vec 0}_\parallel \equiv \vec d^{\vec 0} \cdot \vec e^{\vec 0}_\parallel$.  In what follows, we consider the case where $d^{\vec 0}_\parallel > 0$.  The case of $d^{\vec 0}_\parallel < 0$ is essentially the same by flipping the coordinate basis.

On the image plane, according to \refeq{photometric-center-x} and \refeq{1d-lens-mapping}, from symmetry, $\vec C_x = \vec 0$.  Then it follows from \refeq{photometric-moment-x} and \refeq{size-x} that
\begin{equation} \label{eq:size-x-explicit}
    \sigma^2_x = \frac{\sigma_\mathrm{S}}{d^{\vec 0}_\parallel}\, f \left(\frac{y_{\mathrm{S}, \parallel}}{\sigma_\mathrm{S}}\right),
\end{equation}
where $f(\xi)$ has the explicit form
\begin{equation}
    f(\xi) \equiv \begin{dcases} 
        \frac{\xi^2 I_{-1/4}(\xi^2/4) + (\xi^2 + 2) I_{1/4}(\xi^2 / 4) + \xi^2 [I_{3/4}(\xi^2/4) + I_{5/4}(\xi^2/4)]}{\xi [I_{-1/4}(\xi^2/4) + I_{1/4}(\xi^2/4)]} & (\xi \geq 0), \\
        \frac{\pi}{\sqrt{2}} \cdot \frac{\xi^2 I_{-1/4}(\xi^2/4) - (\xi^2 + 2) I_{1/4}(\xi^2/4) + \xi^2 [I_{3/4}(\xi^2/4) - I_{5/4}(\xi^2/4)]}{\xi K_{1/4}(\xi^2/4)} & (\xi < 0),
    \end{dcases}
\end{equation}
with $I_\nu(z)$ and $K_\nu(z)$ being the modified Bessel function of the first and the second kind, respectively.  On the source plane, the visible side of the fold is the side in which $y_\parallel \geq 0$.  So according to \refeq{photometric-center-visible-y}, we have
\begin{equation}
    \vec C_{\mathrm{V}, y} = \left(y_{\mathrm{S}, \parallel} + \sigma_\mathrm{S} \sqrt{\frac{2}{\pi}} \frac{\exp \left[- y_{\mathrm{S}, \parallel}^2 / (2 \sigma_\mathrm{S}^2) \right]}{1 + \erf \left[y_{\mathrm{S}, \parallel}/(\sqrt2 \sigma_\mathrm{S}) \right]} \right) \vec e^{\vec 0}_\parallel.
\end{equation}
Then it follows from \refeq{photometric-moment-visible-y} and \refeq{size-visible-y} that
\begin{equation} \label{eq:size-visible-y-explicit}
    \sigma_\mathrm{V}^2 = \sigma_\mathrm{S}^2\, g \left( \frac{y_{\mathrm{S}, \parallel}}{\sigma_\mathrm{S}} \right),
\end{equation}
where $g(\xi)$ has the explicit form
\begin{equation}
    g(\xi) \equiv 1 - \xi \sqrt{\frac{2}{\pi}}\frac{\exp(-\xi^2/2)}{1 + \erf(\xi / \sqrt2)} - \frac{2}{\pi} \left[\frac{\exp(-\xi^2/2)}{1 + \erf(\xi / \sqrt2)}\right]^2.
\end{equation}

\begin{figure}
    \centering
    \includegraphics[width = \linewidth]{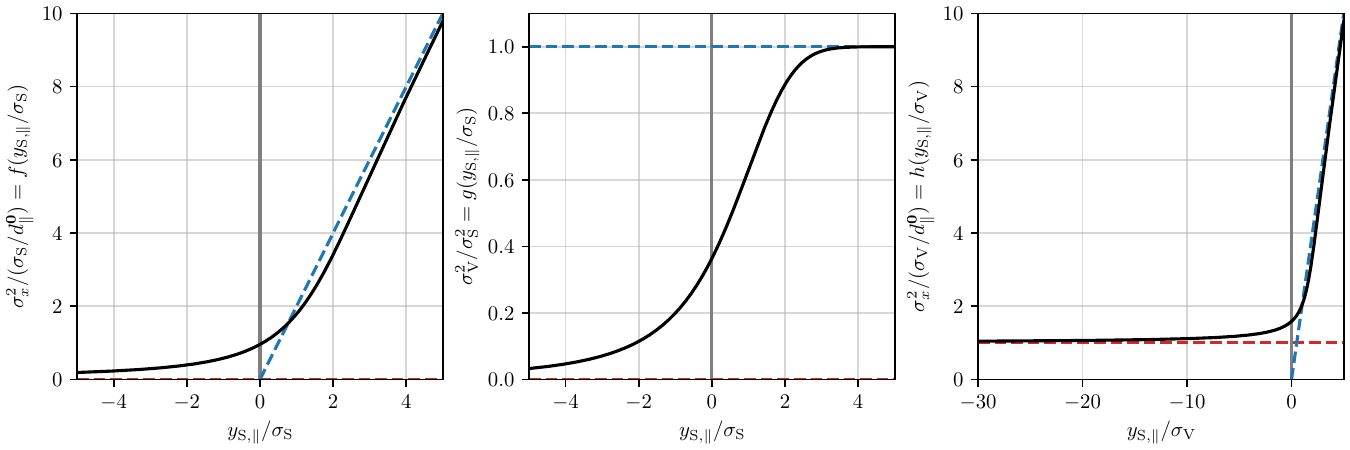}
    \caption{Functions $f(\xi)$, $g(\xi)$, and $h(\eta)$ (from left to right) as defined in \refeq{size-x-explicit}, \refeq{size-visible-y-explicit}, and \refeq{size-x-from-size-visible-y}, respectively.  The left and right asymptotes of each function are shown in red and blue dashed lines, respectively.}
    \label{fig:functions-f-g-h}
\end{figure}

\begin{figure}
    \centering
    \includegraphics[width = \linewidth]{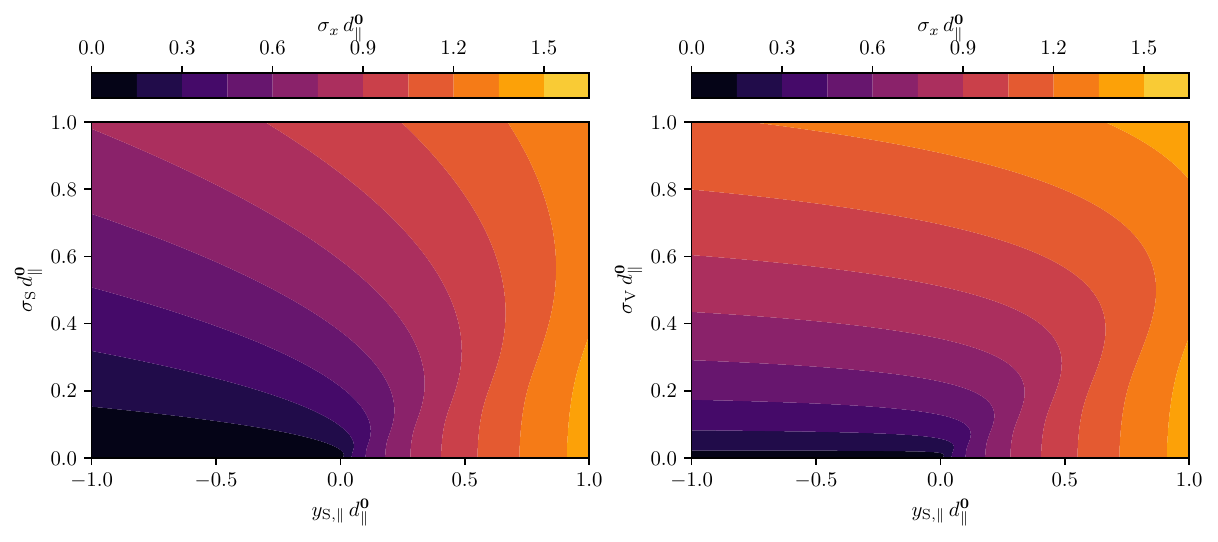}
    \caption{$\sigma_x$ as a function of $y_{\mathrm S, \parallel}$ and $\sigma_\mathrm{S}$ (left) or $\sigma_\mathrm{V}$ (right).  All quantities are in units of $1 / d^{\vec 0}_\parallel$.  For a given resolution $\mathcal R$ (i.e.\ a given contour line in the plot), when $\sigma_\mathrm{S}$ is used to quote the size constraints, a source, no matter how big, can always appear unresolved by having a very negative $y_{\mathrm S, \parallel}$.  This is not the case for $\sigma_\mathrm{V}$ as the contours flatten out for very negative $y_{\mathrm S, \parallel}$.  So a valid size constraint can be derived using $\sigma_\mathrm{V}$.}
    \label{fig:subfold-sigma-x}
\end{figure}

We now express $\sigma_x$ in terms of $\sigma_\mathrm{V}$ by combining \refeq{size-x-explicit} and \refeq{size-visible-y-explicit} to eliminate $\sigma_\mathrm{S}$.  Through dimensional analysis, the result must take the form
\begin{equation} \label{eq:size-x-from-size-visible-y}
    \sigma^2_x = \frac{\sigma_\mathrm{V}}{d^{\vec 0}_\parallel}\, h \left(\frac{y_{\mathrm{S}, \parallel}}{\sigma_\mathrm{V}}\right),
\end{equation}
where $h(\eta)$ is the dimensionless function to be determined.  By using \refeq{size-visible-y-explicit}, \refeq{size-x-from-size-visible-y} can be cast as
\begin{equation}
    \sigma^2_x 
    = \frac{\sigma_\mathrm{V}}{d^{\vec 0}_\parallel}\, h \left(\frac{y_{\mathrm{S}, \parallel}}{\sigma_\mathrm{V}}\right) 
    = \frac{\sigma_\mathrm{S}}{d^{\vec 0}_\parallel} \frac{\sigma_\mathrm{V}}{\sigma_\mathrm{S}}\, h \left(\frac{y_{\mathrm{S}, \parallel}}{\sigma_\mathrm{S}} \frac{\sigma_\mathrm{S}}{\sigma_\mathrm{V}}\right)
    = \frac{\sigma_\mathrm{S}}{d^{\vec 0}_\parallel} \left[g \left( \frac{y_{\mathrm{S}, \parallel}}{\sigma_\mathrm{S}} \right)\right]^{\frac{1}{2}}\, h \left(\frac{y_{\mathrm{S}, \parallel}}{\sigma_\mathrm{S}} \left[g \left( \frac{y_{\mathrm{S}, \parallel}}{\sigma_\mathrm{S}} \right)\right]^{-\frac{1}{2}} \right).
\end{equation}
We compare this with \refeq{size-x-explicit}, finding that function $h(\eta)$ is implicitly, but uniquely, defined by
\begin{equation}
    h\left(\frac{\xi}{[g(\xi)]^{1/2}}\right) = \frac{f(\xi)}{[g(\xi)]^{1/2}}.
\end{equation}
We are not able to find an explicit formula for $h(\eta)$.  The functions $f(\xi)$, $g(\xi)$, and $h(\eta)$ are plotted in \reffig{functions-f-g-h}.  The fact that $h(\eta)$ has range $(1,+\infty)$, together with \refeq{size-x-from-size-visible-y} and \refeq{size-visible-y-constraint}, implies that $\overline \sigma_\mathrm{V} = \mathcal R^2 d_\parallel^{\vec 0}$, which then implies \refeq{physical-size-constraint}.  The reason why using $\sigma_\mathrm{V}$ in place of $\sigma_\mathrm{S}$ will remove the corner case mentioned in \refsec{size-constraints} is illustrated in \reffig{subfold-sigma-x}.

\bibliography{references}{}

\begin{thebibliography}{}
\expandafter\ifx\csname natexlab\endcsname\relax\def\natexlab#1{#1}\fi
\providecommand{\url}[1]{\href{#1}{#1}}
\providecommand{\dodoi}[1]{doi:~\href{http://doi.org/#1}{\nolinkurl{#1}}}
\providecommand{\doeprint}[1]{\href{http://ascl.net/#1}{\nolinkurl{http://ascl.net/#1}}}
\providecommand{\doarXiv}[1]{\href{https://arxiv.org/abs/#1}{\nolinkurl{https://arxiv.org/abs/#1}}}

\bibitem[{{Amruth} {et~al.}(2023){Amruth}, {Broadhurst}, {Lim}, {Oguri},
  {Smoot}, {Diego}, {Leung}, {Emami}, {Li}, {Chiueh}, {Schive}, {Yeung}, \&
  {Li}}]{Amruth2023EinsteinRingWaveDM}
{Amruth}, A., {Broadhurst}, T., {Lim}, J., {et~al.} 2023, Nature Astronomy, 7,
  736, \dodoi{10.1038/s41550-023-01943-9}

\bibitem[{{Baltz} {et~al.}(2009){Baltz}, {Marshall}, \&
  {Oguri}}]{Baltz2009AnalyticLensModels}
{Baltz}, E.~A., {Marshall}, P., \& {Oguri}, M. 2009, \jcap, 2009, 015,
  \dodoi{10.1088/1475-7516/2009/01/015}

\bibitem[{{Barmby} {et~al.}(2007){Barmby}, {McLaughlin}, {Harris}, {Harris}, \&
  {Forbes}}]{Barmby2007M31GCcatalog}
{Barmby}, P., {McLaughlin}, D.~E., {Harris}, W.~E., {Harris}, G. L.~H., \&
  {Forbes}, D.~A. 2007, \aj, 133, 2764, \dodoi{10.1086/516777}

\bibitem[{{Bartels} \& {Ando}(2015)}]{2015PhRvD..92l3508B}
{Bartels}, R., \& {Ando}, S. 2015, \prd, 92, 123508,
  \dodoi{10.1103/PhysRevD.92.123508}

\bibitem[{{Baumgardt} \& {Hilker}(2018)}]{Baumgardt2018MWGCcatalog}
{Baumgardt}, H., \& {Hilker}, M. 2018, \mnras, 478, 1520,
  \dodoi{10.1093/mnras/sty1057}

\bibitem[{{Broadhurst} {et~al.}(2024){Broadhurst}, {Li}, {Alfred}, {Diego},
  {Morilla}, {Kelly}, {Sun}, {Oguri}, {Williams}, {Windhorst}, {Zitrin}, {Abe},
  {Chen}, {Fudamoto}, {Kawai}, {Lim}, {Liu}, {Meena}, {Palencia}, {Smoot}, \&
  {Williams}}]{Broadhurst2024arXiv240519422B}
{Broadhurst}, T., {Li}, S.~K., {Alfred}, A., {et~al.} 2024, arXiv e-prints,
  arXiv:2405.19422, \dodoi{10.48550/arXiv.2405.19422}

\bibitem[{{Bullock} {et~al.}(2001){Bullock}, {Kolatt}, {Sigad}, {Somerville},
  {Kravtsov}, {Klypin}, {Primack}, \& {Dekel}}]{2001MNRAS.321..559B}
{Bullock}, J.~S., {Kolatt}, T.~S., {Sigad}, Y., {et~al.} 2001, \mnras, 321,
  559, \dodoi{10.1046/j.1365-8711.2001.04068.x}

\bibitem[{{Chen} {et~al.}(2019){Chen}, {Kelly}, {Diego}, {Oguri}, {Williams},
  {Zitrin}, {Treu}, {Smith}, {Broadhurst}, {Kaiser}, {Foley}, {Filippenko},
  {Salo}, {Hjorth}, \& {Selsing}}]{Chen2019LensedStarM0416}
{Chen}, W., {Kelly}, P.~L., {Diego}, J.~M., {et~al.} 2019, \apj, 881, 8,
  \dodoi{10.3847/1538-4357/ab297d}

\bibitem[{{Choe} {et~al.}(2024){Choe}, {Rivera-Thorsen}, {Dahle}, {Sharon},
  {Owens}, {Rigby}, {Bayliss}, {Hayes}, {Hutchison}, {Welch}, {Chisholm},
  {Gladders}, {Khullar}, \& {Kim}}]{Choe2024Godzilla}
{Choe}, S., {Rivera-Thorsen}, T.~E., {Dahle}, H., {et~al.} 2024, arXiv
  e-prints, arXiv:2405.06953, \dodoi{10.48550/arXiv.2405.06953}

\bibitem[{{Cyr-Racine} {et~al.}(2016){Cyr-Racine}, {Moustakas}, {Keeton},
  {Sigurdson}, \& {Gilman}}]{CyrRacine2016DarkCensus}
{Cyr-Racine}, F.-Y., {Moustakas}, L.~A., {Keeton}, C.~R., {Sigurdson}, K., \&
  {Gilman}, D.~A. 2016, \prd, 94, 043505, \dodoi{10.1103/PhysRevD.94.043505}

\bibitem[{{Dai}(2021)}]{Dai2021StarClusterMicrolensing}
{Dai}, L. 2021, \mnras, 501, 5538, \dodoi{10.1093/mnras/stab017}

\bibitem[{{Dai} \& {Miralda-Escud{\'e}}(2020)}]{Dai2020QCDAxionMinihalos}
{Dai}, L., \& {Miralda-Escud{\'e}}, J. 2020, \aj, 159, 49,
  \dodoi{10.3847/1538-3881/ab5e83}

\bibitem[{{Dai} \& {Pascale}(2021)}]{Dai2021NewApproxMagnificationStatistics}
{Dai}, L., \& {Pascale}, M. 2021, arXiv e-prints, arXiv:2104.12009,
  \dodoi{10.48550/arXiv.2104.12009}

\bibitem[{{Dai} {et~al.}(2018){Dai}, {Venumadhav}, {Kaurov}, \&
  {Miralda-Escud}}]{Dai2018Abell370}
{Dai}, L., {Venumadhav}, T., {Kaurov}, A.~A., \& {Miralda-Escud}, J. 2018,
  \apj, 867, 24, \dodoi{10.3847/1538-4357/aae478}

\bibitem[{{Dai} {et~al.}(2020){Dai}, {Kaurov}, {Sharon}, {Florian},
  {Miralda-Escud{\'e}}, {Venumadhav}, {Frye}, {Rigby}, \&
  {Bayliss}}]{Dai2020SGASJ1226}
{Dai}, L., {Kaurov}, A.~A., {Sharon}, K., {et~al.} 2020, \mnras, 495, 3192,
  \dodoi{10.1093/mnras/staa1355}

\bibitem[{{Dalal} \& {Kravtsov}(2022)}]{Dalal2022FuzzyDMbounds}
{Dalal}, N., \& {Kravtsov}, A. 2022, \prd, 106, 063517,
  \dodoi{10.1103/PhysRevD.106.063517}

\bibitem[{{Diego} {et~al.}(2022){Diego}, {Pascale}, {Kavanagh}, {Kelly}, {Dai},
  {Frye}, \& {Broadhurst}}]{Diego2022Godzilla}
{Diego}, J.~M., {Pascale}, M., {Kavanagh}, B.~J., {et~al.} 2022, \aap, 665,
  A134, \dodoi{10.1051/0004-6361/202243605}

\bibitem[{{Diego} {et~al.}(2018){Diego}, {Kaiser}, {Broadhurst}, {Kelly},
  {Rodney}, {Morishita}, {Oguri}, {Ross}, {Zitrin}, {Jauzac}, {Richard},
  {Williams}, {Vega-Ferrero}, {Frye}, \&
  {Filippenko}}]{Diego2018DMUnderMicroscope}
{Diego}, J.~M., {Kaiser}, N., {Broadhurst}, T., {et~al.} 2018, \apj, 857, 25,
  \dodoi{10.3847/1538-4357/aab617}

\bibitem[{{Diego} {et~al.}(2024{\natexlab{a}}){Diego}, {Li}, {Amruth}, {Meena},
  {Broadhurst}, {Kelly}, {Filippenko}, {Williams}, {Zitrin}, {Harris},
  {Reina-Campos}, {Giocoli}, {Dai}, {Struble}, {Treu}, {Fudamoto}, {Gilman},
  {Koekemoer}, {Lim}, {Palencia}, {Sun}, \&
  {Windhorst}}]{Diego2024arXiv240408033D}
{Diego}, J.~M., {Li}, S.~K., {Amruth}, A., {et~al.} 2024{\natexlab{a}}, arXiv
  e-prints, arXiv:2404.08033, \dodoi{10.48550/arXiv.2404.08033}

\bibitem[{{Diego} {et~al.}(2024{\natexlab{b}}){Diego}, {Amruth}, {Palencia},
  {Broadhurst}, {Li}, {Lim}, {Windhorst}, {Zitrin}, {Filippenko}, {Williams},
  {Meena}, {Chen}, \& {Kelly}}]{Diego2024arXiv240608537D}
{Diego}, J.~M., {Amruth}, A., {Palencia}, J.~M., {et~al.} 2024{\natexlab{b}},
  arXiv e-prints, arXiv:2406.08537, \dodoi{10.48550/arXiv.2406.08537}

\bibitem[{{Diemand} {et~al.}(2007){Diemand}, {Kuhlen}, \&
  {Madau}}]{2007ApJ...667..859D}
{Diemand}, J., {Kuhlen}, M., \& {Madau}, P. 2007, \apj, 667, 859,
  \dodoi{10.1086/520573}

\bibitem[{{Diemer} \& {Joyce}(2019)}]{2019ApJ...871..168D}
{Diemer}, B., \& {Joyce}, M. 2019, \apj, 871, 168,
  \dodoi{10.3847/1538-4357/aafad6}

\bibitem[{{Fudamoto} {et~al.}(2024){Fudamoto}, {Sun}, {Diego}, {Dai}, {Oguri},
  {Zitrin}, {Zackrisson}, {Jauzac}, {Lagattuta}, {Egami}, {Iani}, {Windhorst},
  {Abe}, {Bauer}, {Bian}, {Bhatawdekar}, {Broadhurst}, {Cai}, {Chen}, {Chen},
  {Cohen}, {Conselice}, {Espada}, {Foo}, {Frye}, {Fujimoto}, {Furtak},
  {Golubchik}, {Hsiao}, {Jolly}, {Kawai}, {Kelly}, {Koekemoer}, {Kohno},
  {Kokorev}, {Li}, {Li}, {Lin}, {Magdis}, {Meena}, {Nabizadeh}, {Richard},
  {Steinhardt}, {Wu}, {Zhu}, \& {Zou}}]{Fudamoto2024Abell370Dragon}
{Fudamoto}, Y., {Sun}, F., {Diego}, J.~M., {et~al.} 2024, arXiv e-prints,
  arXiv:2404.08045, \dodoi{10.48550/arXiv.2404.08045}

\bibitem[{{Furtak} {et~al.}(2023){Furtak}, {Meena}, {Zackrisson}, {Zitrin},
  {Brammer}, {Coe}, {Diego}, {Eldridge}, {Jim{\'e}nez-Teja}, {Kokorev},
  {Ricotti}, {Welch}, {Windhorst}, {Abdurro'uf}, {Andrade-Santos},
  {Bhatawdekar}, {Bradley}, {Broadhurst}, {Chen}, {Conselice}, {Dayal}, {Frye},
  {Fujimoto}, {Hsiao}, {Kelly}, {Mahler}, {Mandelker}, {Norman}, {Oguri},
  {Pirzkal}, {Postman}, {Ravindranath}, {Vanzella}, \&
  {Wilkins}}]{Furtak2023arXiv230800042F}
{Furtak}, L.~J., {Meena}, A.~K., {Zackrisson}, E., {et~al.} 2023, arXiv
  e-prints, arXiv:2308.00042, \dodoi{10.48550/arXiv.2308.00042}

\bibitem[{Han {et~al.}(2016)Han, Cole, Frenk, \& Jing}]{Han:2015pua}
Han, J., Cole, S., Frenk, C.~S., \& Jing, Y. 2016, Mon. Not. Roy. Astron. Soc.,
  457, 1208, \dodoi{10.1093/mnras/stv2900}

\bibitem[{{Han} \& {Dai}(2024)}]{Han2024HighlyMagnifiedStar}
{Han}, X., \& {Dai}, L. 2024, \apj, 964, 160, \dodoi{10.3847/1538-4357/ad2b6a}

\bibitem[{{Hunter} {et~al.}(1995){Hunter}, {Shaya}, {Holtzman}, {Light},
  {O'Neil}, \& {Lynds}}]{Hunter1995R136HSTStarCount}
{Hunter}, D.~A., {Shaya}, E.~J., {Holtzman}, J.~A., {et~al.} 1995, \apj, 448,
  179, \dodoi{10.1086/175950}

\bibitem[{{Kaurov} {et~al.}(2019){Kaurov}, {Dai}, {Venumadhav},
  {Miralda-Escud{\'e}}, \& {Frye}}]{Kaurov2019LensedStarM0416}
{Kaurov}, A.~A., {Dai}, L., {Venumadhav}, T., {Miralda-Escud{\'e}}, J., \&
  {Frye}, B. 2019, \apj, 880, 58, \dodoi{10.3847/1538-4357/ab2888}

\bibitem[{{Kelly} {et~al.}(2018){Kelly}, {Diego}, {Rodney}, {Kaiser},
  {Broadhurst}, {Zitrin}, {Treu}, {P{\'e}rez-Gonz{\'a}lez}, {Morishita},
  {Jauzac}, {Selsing}, {Oguri}, {Pueyo}, {Ross}, {Filippenko}, {Smith},
  {Hjorth}, {Cenko}, {Wang}, {Howell}, {Richard}, {Frye}, {Jha}, {Foley},
  {Norman}, {Bradac}, {Zheng}, {Brammer}, {Benito}, {Cava}, {Christensen}, {de
  Mink}, {Graur}, {Grillo}, {Kawamata}, {Kneib}, {Matheson}, {McCully},
  {Nonino}, {P{\'e}rez-Fournon}, {Riess}, {Rosati}, {Schmidt}, {Sharon}, \&
  {Weiner}}]{Kelly2018NatAsM1149}
{Kelly}, P.~L., {Diego}, J.~M., {Rodney}, S., {et~al.} 2018, Nature Astronomy,
  2, 334, \dodoi{10.1038/s41550-018-0430-3}

\bibitem[{{Kelly} {et~al.}(2022){Kelly}, {Chen}, {Alfred}, {Broadhurst},
  {Diego}, {Emami}, {Filippenko}, {Keen}, {Kei Li}, {Lim}, {Meena}, {Oguri},
  {Scarlata}, {Treu}, {Williams}, {Williams}, {Zhou}, {Zitrin}, {Foley}, {Jha},
  {Kaiser}, {Mehta}, {Rieck}, {Salo}, {Smith}, \&
  {Weisz}}]{Kelly2022FlashlightDozenStars}
{Kelly}, P.~L., {Chen}, W., {Alfred}, A., {et~al.} 2022, arXiv e-prints,
  arXiv:2211.02670, \dodoi{10.48550/arXiv.2211.02670}

\bibitem[{{Krumholz} {et~al.}(2019){Krumholz}, {McKee}, \&
  {Bland-Hawthorn}}]{KMBH2019ARAAreview}
{Krumholz}, M.~R., {McKee}, C.~F., \& {Bland-Hawthorn}, J. 2019, \araa, 57,
  227, \dodoi{10.1146/annurev-astro-091918-104430}

\bibitem[{Lin {et~al.}(2022)Lin, Wagner, \& Griffiths}]{Lin:2022mrn}
Lin, J., Wagner, J., \& Griffiths, R.~E. 2022, Mon. Not. Roy. Astron. Soc.,
  517, 1821, \dodoi{10.1093/mnras/stac2576}

\bibitem[{{Ludlow} {et~al.}(2016){Ludlow}, {Bose}, {Angulo}, {Wang},
  {Hellwing}, {Navarro}, {Cole}, \& {Frenk}}]{2016MNRAS.460.1214L}
{Ludlow}, A.~D., {Bose}, S., {Angulo}, R.~E., {et~al.} 2016, \mnras, 460, 1214,
  \dodoi{10.1093/mnras/stw1046}

\bibitem[{{Lundqvist} {et~al.}(2024){Lundqvist}, {Zackrisson}, {Hawcroft},
  {Amarsi}, \& {Welch}}]{Lundqvist2024arXiv240410817L}
{Lundqvist}, E., {Zackrisson}, E., {Hawcroft}, C., {Amarsi}, A.~M., \& {Welch},
  B. 2024, arXiv e-prints, arXiv:2404.10817, \dodoi{10.48550/arXiv.2404.10817}

\bibitem[{{Marsh} \& {Niemeyer}(2019)}]{Marsh2019FuzzyDMBoundEridanusII}
{Marsh}, D. J.~E., \& {Niemeyer}, J.~C. 2019, \prl, 123, 051103,
  \dodoi{10.1103/PhysRevLett.123.051103}

\bibitem[{{Meena} {et~al.}(2023){Meena}, {Chen}, {Zitrin}, {Kelly},
  {Golubchik}, {Zhou}, {Alfred}, {Broadhurst}, {Diego}, {Filippenko}, {Li},
  {Oguri}, {Smith}, \& {Williams}}]{Meena2023Abell370LensedStar}
{Meena}, A.~K., {Chen}, W., {Zitrin}, A., {et~al.} 2023, \mnras, 521, 5224,
  \dodoi{10.1093/mnras/stad869}

\bibitem[{{Metcalf} \& {Petkova}(2014)}]{2014MNRAS.445.1942M}
{Metcalf}, R.~B., \& {Petkova}, M. 2014, \mnras, 445, 1942,
  \dodoi{10.1093/mnras/stu1859}

\bibitem[{{Miralda-Escude}(1991)}]{Miralda1991CausticCrossingStars}
{Miralda-Escude}, J. 1991, \apj, 379, 94, \dodoi{10.1086/170486}

\bibitem[{{Molin{\'e}} {et~al.}(2017){Molin{\'e}}, {S{\'a}nchez-Conde},
  {Palomares-Ruiz}, \& {Prada}}]{2017MNRAS.466.4974M}
{Molin{\'e}}, {\'A}., {S{\'a}nchez-Conde}, M.~A., {Palomares-Ruiz}, S., \&
  {Prada}, F. 2017, \mnras, 466, 4974, \dodoi{10.1093/mnras/stx026}

\bibitem[{{Nabizadeh} {et~al.}(2024){Nabizadeh}, {Zackrisson}, {Lundqvist},
  {Ricotti}, {Park}, {Welch}, \& {Diego}}]{Nabizadeh2024arXiv240612607N}
{Nabizadeh}, A., {Zackrisson}, E., {Lundqvist}, E., {et~al.} 2024, arXiv
  e-prints, arXiv:2406.12607, \dodoi{10.48550/arXiv.2406.12607}

\bibitem[{{Navarro} {et~al.}(1996){Navarro}, {Frenk}, \&
  {White}}]{Navarro1996NFWprofile}
{Navarro}, J.~F., {Frenk}, C.~S., \& {White}, S.~D.~M. 1996, \apj, 462, 563,
  \dodoi{10.1086/177173}

\bibitem[{{Oguri} {et~al.}(2018){Oguri}, {Diego}, {Kaiser}, {Kelly}, \&
  {Broadhurst}}]{Oguri2018CausticMicrolensing}
{Oguri}, M., {Diego}, J.~M., {Kaiser}, N., {Kelly}, P.~L., \& {Broadhurst}, T.
  2018, \prd, 97, 023518, \dodoi{10.1103/PhysRevD.97.023518}

\bibitem[{{Okamoto} {et~al.}(2008){Okamoto}, {Gao}, \&
  {Theuns}}]{Okamoto2008StarFormationInefficiencySmallGalaxies}
{Okamoto}, T., {Gao}, L., \& {Theuns}, T. 2008, \mnras, 390, 920,
  \dodoi{10.1111/j.1365-2966.2008.13830.x}

\bibitem[{{Pascale} \& {Dai}(2024)}]{Pascale2024Godzilla}
{Pascale}, M., \& {Dai}, L. 2024, arXiv e-prints, arXiv:2404.10755,
  \dodoi{10.48550/arXiv.2404.10755}

\bibitem[{{Press} \& {Schechter}(1974)}]{Press1974PressSchechter}
{Press}, W.~H., \& {Schechter}, P. 1974, \apj, 187, 425, \dodoi{10.1086/152650}

\bibitem[{{Rodney} {et~al.}(2018){Rodney}, {Balestra}, {Bradac}, {Brammer},
  {Broadhurst}, {Caminha}, {Chiriv{\i}}, {Diego}, {Filippenko}, {Foley},
  {Graur}, {Grillo}, {Hemmati}, {Hjorth}, {Hoag}, {Jauzac}, {Jha}, {Kawamata},
  {Kelly}, {McCully}, {Mobasher}, {Molino}, {Oguri}, {Richard}, {Riess},
  {Rosati}, {Schmidt}, {Selsing}, {Sharon}, {Strolger}, {Suyu}, {Treu},
  {Weiner}, {Williams}, \& {Zitrin}}]{Rodney2018NatAsM0416Transients}
{Rodney}, S.~A., {Balestra}, I., {Bradac}, M., {et~al.} 2018, Nature Astronomy,
  2, 324, \dodoi{10.1038/s41550-018-0405-4}

\bibitem[{{Salmon} {et~al.}(2020){Salmon}, {Coe}, {Bradley}, {Bouwens},
  {Brada{\v{c}}}, {Huang}, {Oesch}, {Stark}, {Sharon}, {Trenti}, {Avila},
  {Ogaz}, {Andrade-Santos}, {Carrasco}, {Cerny}, {Dawson}, {Frye}, {Hoag},
  {Johnson}, {Jones}, {Lam}, {Lovisari}, {Mainali}, {Past}, {Paterno-Mahler},
  {Peterson}, {Riess}, {Rodney}, {Ryan}, {Sendra-Server}, {Strait}, {Strolger},
  {Umetsu}, {Vulcani}, \& {Zitrin}}]{Salmon2020RELICSsurvey}
{Salmon}, B., {Coe}, D., {Bradley}, L., {et~al.} 2020, \apj, 889, 189,
  \dodoi{10.3847/1538-4357/ab5a8b}

\bibitem[{{Schauer} {et~al.}(2022){Schauer}, {Bromm}, {Drory}, \&
  {Boylan-Kolchin}}]{Schauer2022EarendelPopIII}
{Schauer}, A. T.~P., {Bromm}, V., {Drory}, N., \& {Boylan-Kolchin}, M. 2022,
  \apjl, 934, L6, \dodoi{10.3847/2041-8213/ac7f9a}

\bibitem[{{Schneider} {et~al.}(1992{\natexlab{a}}){Schneider}, {Ehlers}, \&
  {Falco}}]{SchneiderEhlersFalco1992textbook}
{Schneider}, P., {Ehlers}, J., \& {Falco}, E.~E. 1992{\natexlab{a}},
  {Gravitational Lenses}, \dodoi{10.1007/978-3-662-03758-4}

\bibitem[{{Schneider} {et~al.}(1992{\natexlab{b}}){Schneider}, {Ehlers}, \&
  {Falco}}]{1992grle.book.....S}
---. 1992{\natexlab{b}}, {Gravitational Lenses},
  \dodoi{10.1007/978-3-662-03758-4}

\bibitem[{{Vall M{\"u}ller} \&
  {Miralda-Escud{\'e}}(2024)}]{Mueller2024arXiv240316989V}
{Vall M{\"u}ller}, C., \& {Miralda-Escud{\'e}}, J. 2024, arXiv e-prints,
  arXiv:2403.16989, \dodoi{10.48550/arXiv.2403.16989}

\bibitem[{{Vanzella} {et~al.}(2020){Vanzella}, {Meneghetti}, {Pastorello},
  {Calura}, {Sani}, {Cupani}, {Caminha}, {Castellano}, {Rosati}, {D'Odorico},
  {Cristiani}, {Grillo}, {Mercurio}, {Nonino}, {Brammer}, \&
  {Hartman}}]{Vanzella2020Bowen}
{Vanzella}, E., {Meneghetti}, M., {Pastorello}, A., {et~al.} 2020, \mnras, 499,
  L67, \dodoi{10.1093/mnrasl/slaa163}

\bibitem[{{Venumadhav} {et~al.}(2017){Venumadhav}, {Dai}, \&
  {Miralda-Escud{\'e}}}]{Venumadhav2017CausticMicrolensing}
{Venumadhav}, T., {Dai}, L., \& {Miralda-Escud{\'e}}, J. 2017, \apj, 850, 49,
  \dodoi{10.3847/1538-4357/aa9575}

\bibitem[{Wagner(2019)}]{Wagner:2019orn}
Wagner, J. 2019, Universe, 5, 177, \dodoi{10.3390/universe5070177}

\bibitem[{Wagner(2022)}]{Wagner:2022ifl}
---. 2022, Astron. Astrophys., 663, A157, \dodoi{10.1051/0004-6361/202243562}

\bibitem[{{Weisenbach} {et~al.}(2024){Weisenbach}, {Anguita},
  {Miralda-Escud{\'e}}, {Oguri}, {Saha}, \&
  {Schechter}}]{Weisenbach2024MLnearMacroCausticsReview}
{Weisenbach}, L., {Anguita}, T., {Miralda-Escud{\'e}}, J., {et~al.} 2024, arXiv
  e-prints, arXiv:2404.08094, \dodoi{10.48550/arXiv.2404.08094}

\bibitem[{{Welch} {et~al.}(2022{\natexlab{a}}){Welch}, {Coe}, {Diego},
  {Zitrin}, {Zackrisson}, {Dimauro}, {Jim{\'e}nez-Teja}, {Kelly}, {Mahler},
  {Oguri}, {Timmes}, {Windhorst}, {Florian}, {de Mink}, {Avila}, {Anderson},
  {Bradley}, {Sharon}, {Vikaeus}, {McCandliss}, {Brada{\v{c}}}, {Rigby},
  {Frye}, {Toft}, {Strait}, {Trenti}, {Sharma}, {Andrade-Santos}, \&
  {Broadhurst}}]{Welch2022EarendelNature}
{Welch}, B., {Coe}, D., {Diego}, J.~M., {et~al.} 2022{\natexlab{a}}, \nat, 603,
  815, \dodoi{10.1038/s41586-022-04449-y}

\bibitem[{{Welch} {et~al.}(2022{\natexlab{b}}){Welch}, {Coe}, {Zackrisson}, {de
  Mink}, {Ravindranath}, {Anderson}, {Brammer}, {Bradley}, {Yoon}, {Kelly},
  {Diego}, {Windhorst}, {Zitrin}, {Dimauro}, {Jim{\'e}nez-Teja}, {Abdurro'uf},
  {Nonino}, {Acebron}, {Andrade-Santos}, {Avila}, {Bayliss}, {Ben{\'\i}tez},
  {Broadhurst}, {Bhatawdekar}, {Brada{\v{c}}}, {Caminha}, {Chen}, {Eldridge},
  {Farag}, {Florian}, {Frye}, {Fujimoto}, {Gomez}, {Henry}, {Hsiao},
  {Hutchison}, {James}, {Joyce}, {Jung}, {Khullar}, {Larson}, {Mahler},
  {Mandelker}, {McCandliss}, {Morishita}, {Newshore}, {Norman}, {O'Connor},
  {Oesch}, {Oguri}, {Ouchi}, {Postman}, {Rigby}, {Ryan}, {Sharma}, {Sharon},
  {Strait}, {Strolger}, {Timmes}, {Toft}, {Trenti}, {Vanzella}, \&
  {Vikaeus}}]{Welch2022EarendelJWSTimaging}
{Welch}, B., {Coe}, D., {Zackrisson}, E., {et~al.} 2022{\natexlab{b}}, \apjl,
  940, L1, \dodoi{10.3847/2041-8213/ac9d39}

\bibitem[{{Williams} {et~al.}(2023){Williams}, {Kelly}, {Treu}, {Amruth},
  {Diego}, {Kei Li}, {Meena}, {Zitrin}, \&
  {Broadhurst}}]{Williams2023FlashlightsDMSubhalo}
{Williams}, L. L.~R., {Kelly}, P.~L., {Treu}, T., {et~al.} 2023, arXiv
  e-prints, arXiv:2304.06064, \dodoi{10.48550/arXiv.2304.06064}

\bibitem[{{Windhorst} {et~al.}(2018){Windhorst}, {Wyithe}, {Alpaslan},
  {Timmes}, {Andrews}, {Kim}, {Kelly}, {Coe}, {Diego}, {Driver}, \&
  {Dijkstra}}]{Windhorst2018PopIIILensedStars}
{Windhorst}, R.~A., {Wyithe}, S., {Alpaslan}, M., {et~al.} 2018, in American
  Astronomical Society Meeting Abstracts, Vol. 232, American Astronomical
  Society Meeting Abstracts \#232, 325.09

\bibitem[{{Yan} {et~al.}(2023){Yan}, {Ma}, {Sun}, {Wang}, {Kelly}, {Diego},
  {Cohen}, {Windhorst}, {Jansen}, {Grogin}, {Beacom}, {Conselice}, {Driver},
  {Frye}, {Coe}, {Marshall}, {Koekemoer}, {Willmer}, {Robotham}, {D'Silva},
  {Summers}, {Nonino}, {Pirzkal}, {Ryan}, {Ortiz}, {Tompkins}, {Bhatawdekar},
  {Cheng}, {Zitrin}, \& {Willner}}]{Yan2023PEARLStransients}
{Yan}, H., {Ma}, Z., {Sun}, B., {et~al.} 2023, \apjs, 269, 43,
  \dodoi{10.3847/1538-4365/ad0298}

\bibitem[{{Zackrisson} {et~al.}(2023){Zackrisson}, {Hultquist}, {Kordt},
  {Diego}, {Nabizadeh}, {Vikaeus}, {Meena}, {Zitrin}, {Volpato}, {Lundqvist},
  {Welch}, {Costa}, \& {Windhorst}}]{Zackrisson2023PopIII}
{Zackrisson}, E., {Hultquist}, A., {Kordt}, A., {et~al.} 2023, arXiv e-prints,
  arXiv:2312.09289, \dodoi{10.48550/arXiv.2312.09289}

\end{thebibliography}
\bibliographystyle{aasjournal}

\end{document}